\documentclass[prl,showpacs,showkeys,twocolumn,notitlepage,10pt]{revtex4-1}
\usepackage{amsfonts,bbold,amsmath,amssymb,graphicx,epstopdf,verbatim,dsfont,color}
\usepackage[english]{babel}

\def \be{\begin{equation*}}
\def \ee{\end{equation*}}

\begin{document}

\title{Thermal radiation and dissipative phase transition in a BEC with local loss}
\author{Dries Sels and Eugene Demler}

\affiliation{Department of Physics, Harvard University, 17 Oxford st., Cambridge, MA 02138, USA}
\date{\today}

\begin{abstract}
We study the dynamics of an atomic BEC subject to local dissipation in the form of atom losses. We show there is a critical loss rate at which the system undergoes a continuous dissipative phase transition from a homogenous state into a state which contains a sonic horizon. The latter drastically alters the behavior of the system by screening the drain. Dissipation leads to two types of fluctuations. First, fluctuations are generated by particles emitted in the reservoir. Both above and below the critical loss, these result in thermal emission of phonons with a temperature set by the loss rate and the chemical potential. The second type of fluctuation results from scattering on the drain and gives rise to a particular correlation pattern that can be observed in the density-density correlation. Aside from correlations between in an out scattered modes, outgoing particles are correlated with localized modes through a process that is reminiscent of Hawking radiation. Finally, we briefly discuss the dynamics of the system when there are two drains, in which case it is possible to construct a black hole laser.
\end{abstract}
\maketitle

\section*{Introduction}
Ultimately, every quantum system is coupled to an environment. This coupling is detrimental in many systems as it causes quantum system to decohere; washing out many interesting quantum phenomena as they tend to be rather fragile. Great strides have been made in isolating and coherently controlling many-body quantum systems, e.g. cold-atoms~\citep{bloch01,kaufman}, Rydberg atoms~\citep{lukin}, trapped ions~\citep{blatt}, superconducting circuits~\citep{martinis}. 

However, controlled coupling to an environment can also be a resource to prepare complex entangled quantum states~\citep{barreiro1,barreiro2,krauter,diehl} and it can even be used to perform universal quantum computation~\citep{verstraete}.  By combining coherent control with suitably engineered dissipative operations one can achieve more complex dynamics than in closed systems. The dynamics of open systems is governed by an interplay between the intrinsic unitary dynamics of the system and the coupling to the environment. This interplay is absent in the closed system and such competition can thus lead to interesting non-equilibrium stationary states (NESS) that can not exist in equilibrium. These non-equilibrium steady states are fixed points of the dissipative Liouvillian and just like ground states of quantum systems -- who's properties might change drastically by changing a parameter in the Hamiltonian-- the nature of the NESS might depend on a parameter in the Liouvillian.  In that sense, systems can undergo a non-equilibrium quantum phase transition~\cite{dallatorre,marino,baumann}.

To arrive at an interesting state one typically has to drive the system, this drive can be coherent such as pumping a cavity with a laser~\cite{wouters,casteels} or incoherent. The latter is more in the spirit of transport measurements where we connect the system to two or more reservoirs to study steady state transport properties. Such transport measurements, have historically been of great insight in probing quantum phenomena, e.g. quantum Hall or the Kondo effect. Recently significant progress has been made in conducting transport experiments in cold-atomic systems~\citep{krinner,chien,brantut,ott}.

One of the intriguing experimental tools available in non-equilibrium condensates is the possibility to generate a sonic event horizon~\citep{garay,carlos,carusotto01,carusotto02}. These systems provide an analog simulation for semi-classical gravity and emit Hawking radiation~\citep{hawking01,hawking02}, as shown by Unruh~\cite{unruh01}. Spontaneous emission of sound waves in moving atomic BEC's was recently experimentally shown~\citep{Steinhauer01}, after an earlier experiment already in the same system already showed stimulated emission~\citep{Steinhauer02}. The latter is the cold atomic version of the pioneering water tank experiment~\citep{weinfurtner}.  

Arguably, in all of these proposals and experiments the behavior of the waves is fairly well understood and captured well within the hydrodynamic theory. In that respect, questions have been raised about how much we can learn about the mysteries surrounding real black holes~\citep{castelvecchi}. In this work we show the importance of UV modes, providing insight into the trans-planckian problem~\citep{jacobson} and the information paradox~\citep{hawking03}.

 \begin{figure}[h]
\includegraphics[width=\columnwidth]{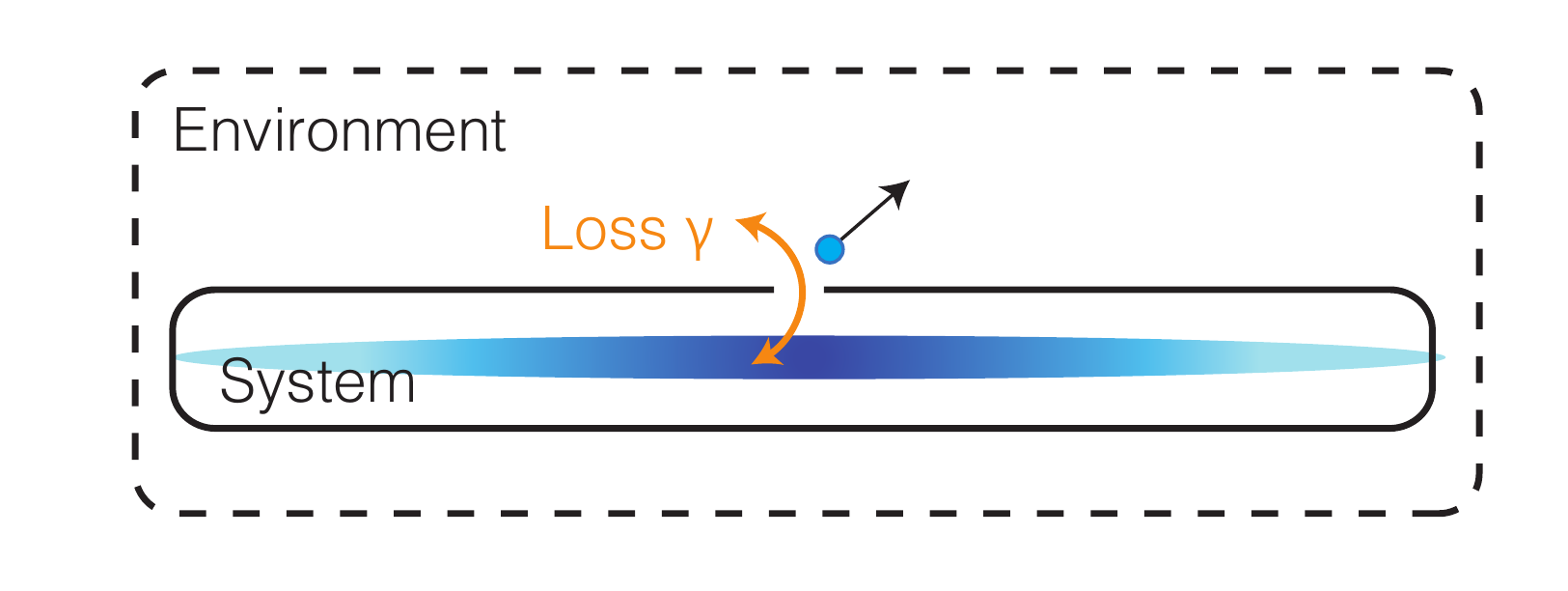}
\caption{{\bf Setup} Atoms are confined in a quasi-one-dimensional tube but can leave the system at the drain with rate $\gamma$. We will consider the system and enviroment to be in the ground state before we suddenly switch on the loss. }
\label{fig:setting}
\end{figure}

We study the formation of a NESS in a BEC subjected to local losses, see Fig.~\ref{fig:setting}. Because the loss is local, the system serves as its own drive and a non-trivial steady state is established in the thermodynamic limit. We will show that--through an interplay between condensate depletion and superfluid flow-- a \emph{Planck size black hole} is formed above a critical dissipation strength. While the hydrodynamic gravitiational analogue correctly identifies the onset of Hawking radiation, it fails to correctly capture the full physics. Unlike in quantum gravity, the UV completion of the problem is known and a microscopic calculation reveals the buildup of correlations between localized and propagating excitations.

\section*{Non-equilibrium steady state}
Consider an elongated BEC in which the transverse confinement is tight enough so that transverse modes are completely gapped out. The effectively one-dimensional system is therefore well described by the Hamiltonian:
\begin{equation*}
H=\int {\rm d}x \left[\frac{\hbar^2}{2m} \partial_x \psi^\dagger(x)\partial_x \psi(x)+\frac{g}{2} \psi^\dagger(x)\psi^\dagger(x) \psi(x)\psi(x)\right], 
\end{equation*}
where $m$ is the mass of the atoms, $g$ is the 1-D effective interaction strength and $\psi^\dagger(x)$ creates a boson at position $x$. Next we couple the system locally to an ideal zero temperature reservoir, which provides a sink which gives rise to a Markov-like loss process for the atoms. The reduced density matrix $\rho$ of the atomic cloud can be described by the following Lindlbad master equation:
\begin{equation}
\dot{\rho}=-\frac{i}{\hbar} \left[H,\rho\right]+\mathcal{D}(\rho),
\label{eq:Lindblad}
\end{equation}
where the dissipative part is given by
\begin{equation*}
\mathcal{D}(\rho)=\int {\rm d}x \gamma(x) \left(2\psi(x)\rho \psi^\dagger(x)-\left\lbrace
\psi^\dagger(x)\psi(x),\rho \right\rbrace \right).
\end{equation*}
Here $\gamma(x)$ sets the local strength of the atom loss and for most of the paper we'll consider it completely local, i.e. $\gamma(x)=\gamma \delta(x)$. Exact simulation of the dynamics~\eqref{eq:Lindblad} for the reduced density matrix is numerically very challenging. However, for weakly interacting systems the exact dynamics is well captured within the so-called truncated Wigner approximation (TWA)~\cite{steel,polkovnikov,sels}. The method goes beyond the time-dependent Bogoliubov approach, as it incorporates the backreaction of the fluctuations on the condensates. It moreover allows to naturally incorporate both initial quantum fluctuations in atomic cloud as well as fluctuations induced by coupling to an external reservoir. The dynamics of the system is effectively described by a stochastic Gross-Pitaevskii equation (GPE):
\begin{equation}
i \hbar \partial_t \psi =\left( -\frac{\hbar^2}{2m} \partial_x^2 \psi+g |\psi|^2-i\hbar \gamma \delta(x) \right)\psi+\eta(t)\delta(x),
\label{eq:GP_TWA}
\end{equation}
where $\eta$ is Gaussian distributed complex-valued white noise with zero mean $\left<\eta(t) \right>=0$ and covariance $\left< \eta^\ast(t) \eta(s)\right>=\hbar^2\gamma \delta(t-s)$. At $t=0$, before the BEC is coupled to the reservoir, the condensate is homogenous and the system is taken to be in the Bogoliubov ground state. Within TWA this implies:
\begin{equation}
\psi_0(x)=\sqrt{n_0}+\sum_k \left(u_k(x)a_k+v_k(x)a_k^\ast\right),
\label{eq:GP_psi0}
\end{equation}
where $u_k(x)$ and $v_k(x)$ are Bogoliubov eigenvectors and the $a_k$'s are Gaussian distributed complex random numbers with zero mean $\left<a_k\right>=0$ and covariance $\left<a_k^\ast a_q\right>=\delta_{k,q}/2$. Expectation values of any observable can simply be computed by averaging the correspoding Weyl symbol~\cite{polkovnikov} over the classical fields $\psi(x,t)$ obtained by propagating the random initial states~\eqref{eq:GP_psi0} with the stochastic GPE~\eqref{eq:GP_TWA}.

\begin{figure}[h]
\includegraphics[width=\columnwidth]{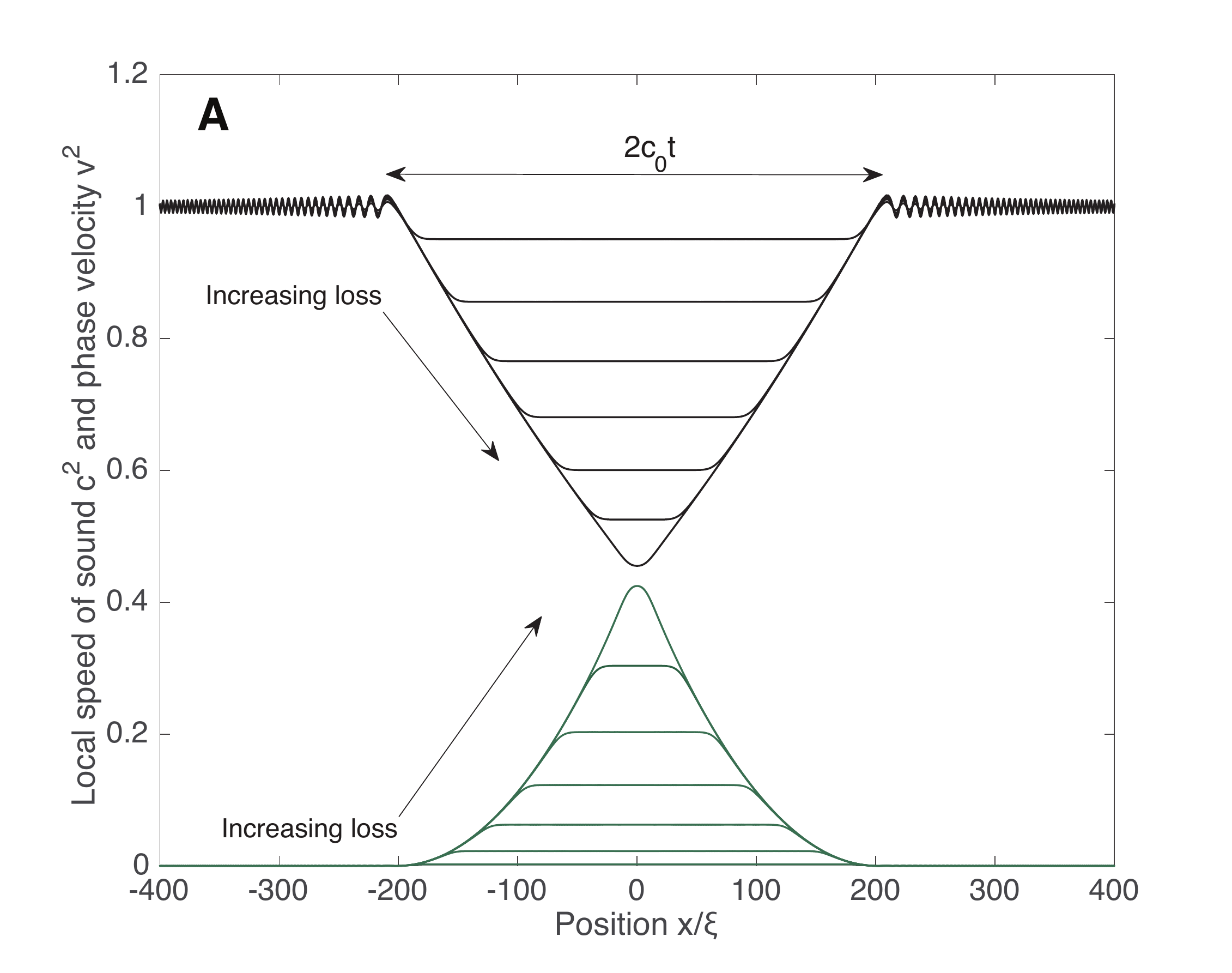}
\includegraphics[width=\columnwidth]{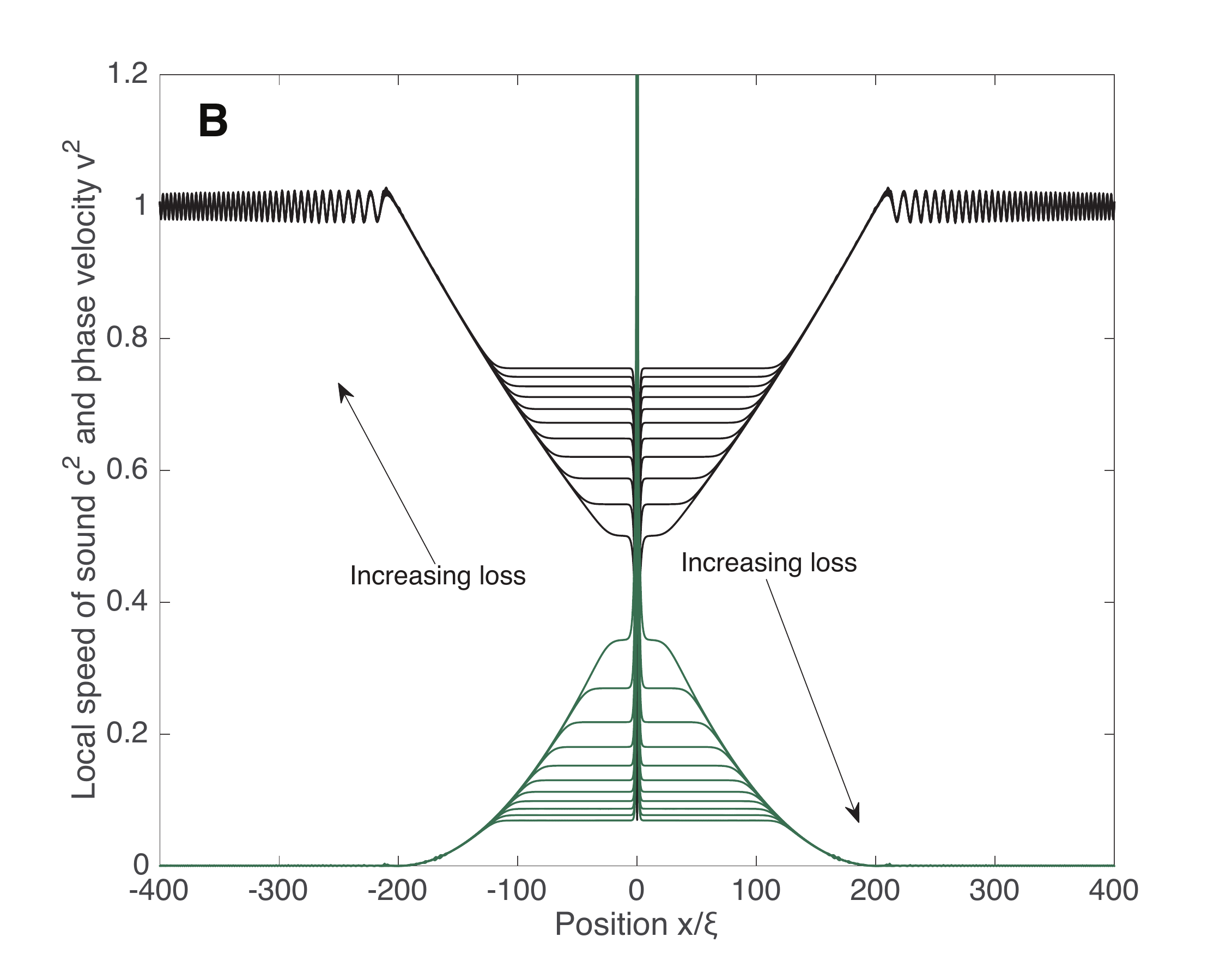}
\caption{{\bf Condensate density and current} The local speed of light $c^2=gn/m$ (in black) and the squared superfluid velocity $v^2=(\partial_x S /m)^2$ (in green). Panel~{\bf A} shows the behavior at low loss rate, i.e. below the critical loss. The condensate gets depleted homogenously within the causal ballistic region. The induced superfluid current moves at a velocity equal to the loss rate $v$. At the critical loss $v^2=c^2$, a horizon is formed around the drain. Panel~{\bf B} shows the behavior above the critical loss. While the velocity keeps increasing \emph{inside} the black hole, it starts to decrease outside the horizon with increasing loss.}
\label{fig:metric}
\end{figure}

\begin{figure}[h]
\includegraphics[width=\columnwidth]{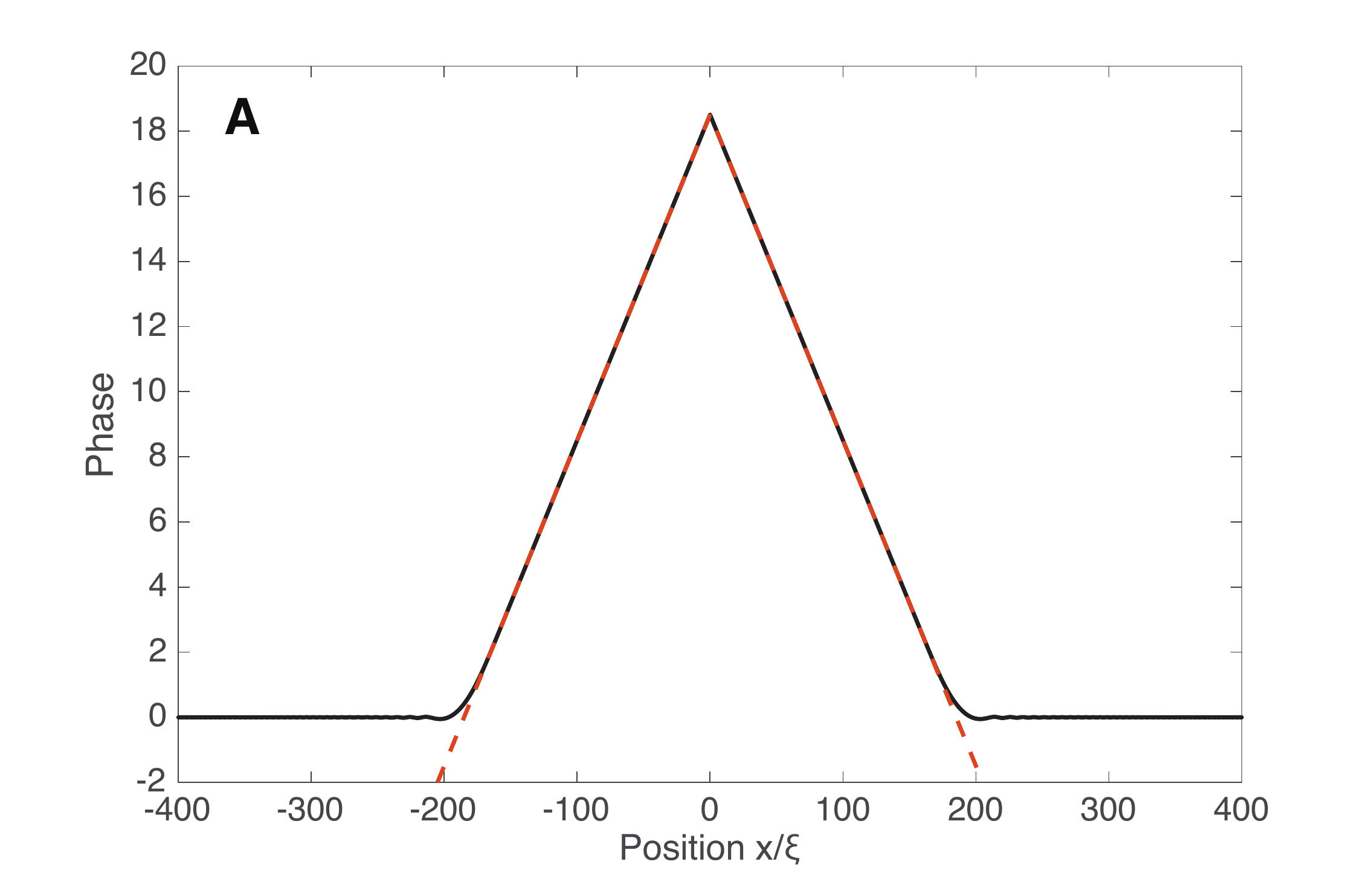}
\includegraphics[width=\columnwidth]{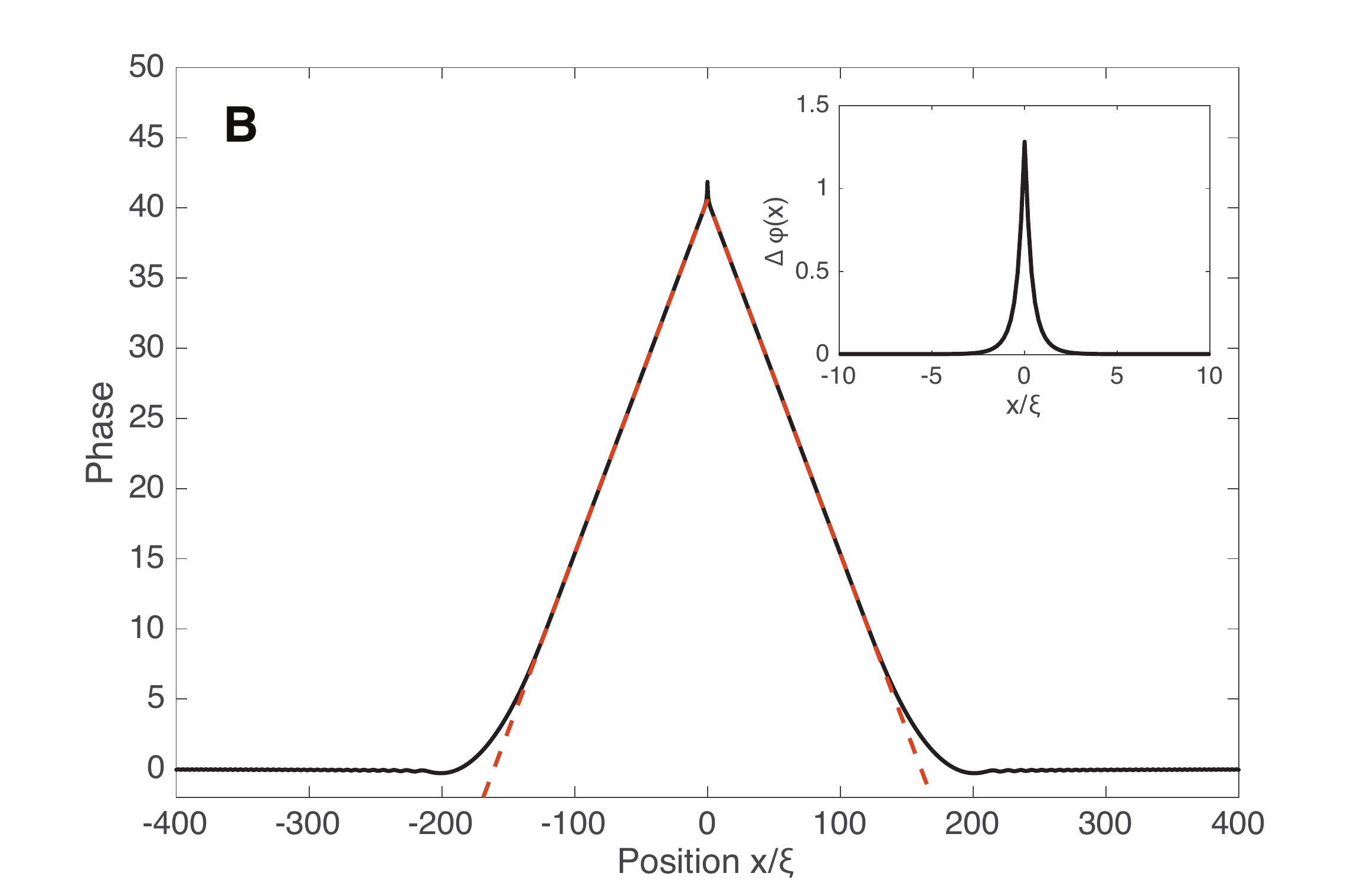}
\caption{{\bf Condensate Phase} The condensate phase at $t=200$ after switching on the loss for a loss rate $\gamma=0.1$ ({\bf A}) and $\gamma=3$ ({\bf B}). Outside the sound cone the phase is zero and inside is almost perfectly described by the stationary solution $S=E t-v|x|$ (red line). At weak coupling $v=\gamma$ while at strong coupling it decreases like $v \propto 1/\gamma$. Above the critical loss, an additional localized peak appears in the phase (see inset in panel {\bf B}).}
\label{fig:phasev01}
\end{figure}

\begin{figure}[h]
\includegraphics[width=\columnwidth]{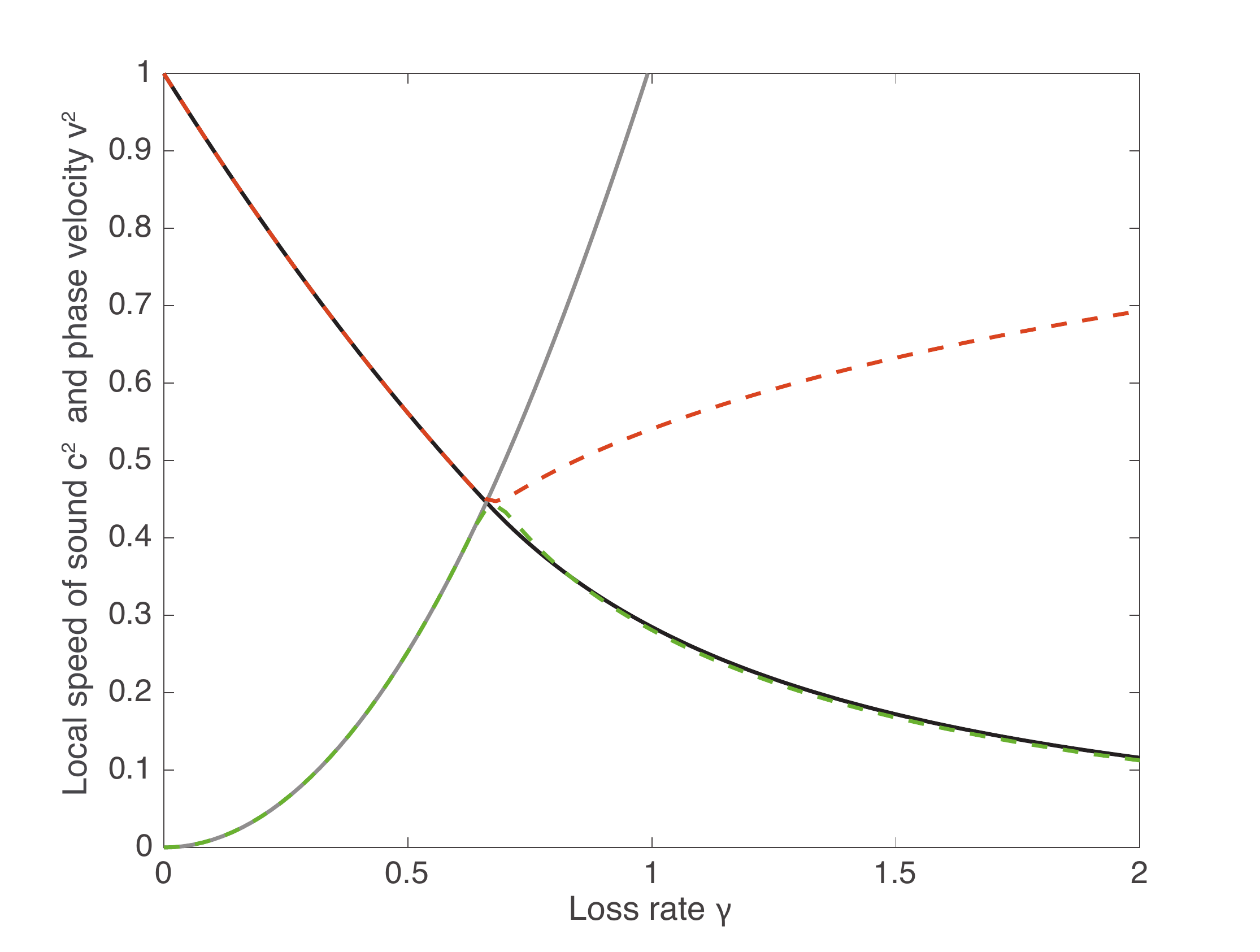}
\caption{{\bf Critical loss} The density at the drain decreases with increasing loss rate (full black line) at the same the velocity at the drain increases (full grey line). At the critical loss rate, the speed of sound at the drain equals the speed of the flow and an horizon is created. Whereas the density (phase) keeps increasing (decreasing) at the drain, outside of the horizon we see a recovery of the density (dashed orange) and phase (dashed green).}
\label{fig:dens_phase}
\end{figure}

Figure~\ref{fig:metric} show the numerical solution of the GPE for various loss rates. After a short transient, which creates a dip in the condensate around the drain, the condensate stops depleting around the drain but the dip ballistically spreads throughout the system. Moreover, for small losses there is a nice linear profile for the phase (see Fig.~\ref{fig:phasev01}) within the causal sound-cone, i.e. there is a homogeneous influx of atoms at velocity $v$ . At the critical loss rate -- when the local condensate flow equals the local speed of sound -- a sonic horizon is formed around the drain. The latter drastically alters the behavior of the system. While the velocity keeps increasing \emph{inside} the black hole, it starts to decrease outside the horizon with increasing loss. Similarly, the density inside the black hole keeps decreasing while it starts to increase again outside. Outside of the black hole, we again see homogenous depletion of the condensate in the causal region (see Fig.~\ref{fig:dens_phase}). Somewhat surprisingly, the phase profile at strong dissipation is remarkably similar to the one at weak dissipation. Outside of the horizon, the phase profile is still linear. Around the drain, a localized peak appears.

To understand the numerics, its instructive to first consider the classical (mean-field) limit, i.e. the evolution of the system in absence of quantum fluctuations. Remarkably, the GPE allows for a non-equilibrium stationary state in thermodynamic limit. Indeed, assuming the density to be homogenous, the condensate wave function can be expressed as $\psi_{S}(x,t)=\sqrt{n}e^{i S(x)/\hbar}e^{-i \mu t/\hbar}$, resulting in the following equations of motion for the phase
\begin{equation}
\partial^2_x S =-2m\gamma\delta(x) \quad {\rm and}\quad \mu=  \frac{(\partial_x S)^2}{2m}+gn.
\label{eq:S1}
\end{equation}
The Green's function of the Laplacian in 1D is just a triangular potenial, consequently we can simulationsly satisfy both equations~\eqref{eq:S1} by:
\begin{equation}
S =-m\gamma |x| \quad {\rm and}\quad \mu=  \frac{m\gamma^2}{2}+gn.
\end{equation}
The chemical potential is thus increased by the kinetic energy in the flow. This clearly explains the behavoir of the condensate below the critical loss. After some short transient, the system goes into a simple quasi-stationary state in which the linear phase profile spread ballistically through the system. Note that the original signal moves in an unperturbed system such that the phase front moves at the bare speed of sound $c_0$. However, after the initial transient the condensate is depleted a bit and a constant inflow with velocity $\gamma$ is established. Consequently, signals only propagate at a velocity $c(x)-\gamma$, where $c(x)$ is the new speed of sound (see Fig.~\ref{fig:dens_phase}). This mismatch in velocities is responsible for the shape of the density profile in panel A of Fig.~\ref{fig:metric}. 

With increasing dissipation, the condensate is further depleted, and at some point the inflow $\gamma$ would exceed the local speed of sound. At that point the system develops an inhomogeneous stationary state. The latter can also be found analytically. The homogenous solution, apart from satisfying eq.~\eqref{eq:S1}, can be thought of a two solutions of the isolated GPE equation suitable matched at the drain. In absence of dissipation, the 1D GPE however also has inhomogeneous integrable solutions. For repulsive interactions those are dark solutions, i.e. $\psi\propto \tanh(x)$.  These solutions have no current, so can not be stationary solutions of the dissipative problem, but grey solutions might be. Taking two grey solutions that flow in opposite direction and matching them in the origin:
\begin{equation}
\psi =\sqrt{n}  \left( i\frac{m v \xi}{\hbar }+\alpha \tanh(\alpha |x|/\xi) \right)e^{-i(mv|x|-\mu t)/\hbar},
\label{eq:soliton}
\end{equation}
with the inverse Lorentz factor and chemical potential 
\begin{eqnarray}
\alpha=\sqrt{1-\frac{v^2}{c^2}}, \quad
\mu=gn\alpha^2+\frac{3 mv^2}{2},
\end{eqnarray}
we indeed find that we can satisfy the GPE equation for $x\neq 0$. Note that this is a solution for any $v$. The cusp in the GPE solution can be used to match the dissipative contribution to the GPE at $x=0$, fixing the velocity to
\begin{equation}
v=\frac{c^2}{\gamma}.
\end{equation}
This is in perfect agreement with the numerics. Above the critical loss rate the velocity outside the horizon is inversely proportional to the dissipation rate while it is exactly linear below the critical loss rate, see~Fig.~\ref{fig:dens_phase}. The stationary flow rate changes continuously with $\gamma$ but has a cusp at the critical loss rate suggesting that $\gamma$ drives the system through a continuous dissipative phase transition. The distinction between weak and strong dissipation is not just some smooth crossover but rather a genuine transition. Recall we gave a microscopic description of why, below the critical loss, it takes longer to develop the stationary state with increasing loss. The latter can now simply be re-interpreted as critical slowing down of the dynamics because we bring the system closer to the critical point.

Since the loss rate controls the flow velocity, which in turn controls the effective red-shift of all the scales we simply get $\nu=1/2$ and since time and space scale in the same way the dynamical exponent $z=1$.

 \begin{figure}[h]
\includegraphics[width=\columnwidth]{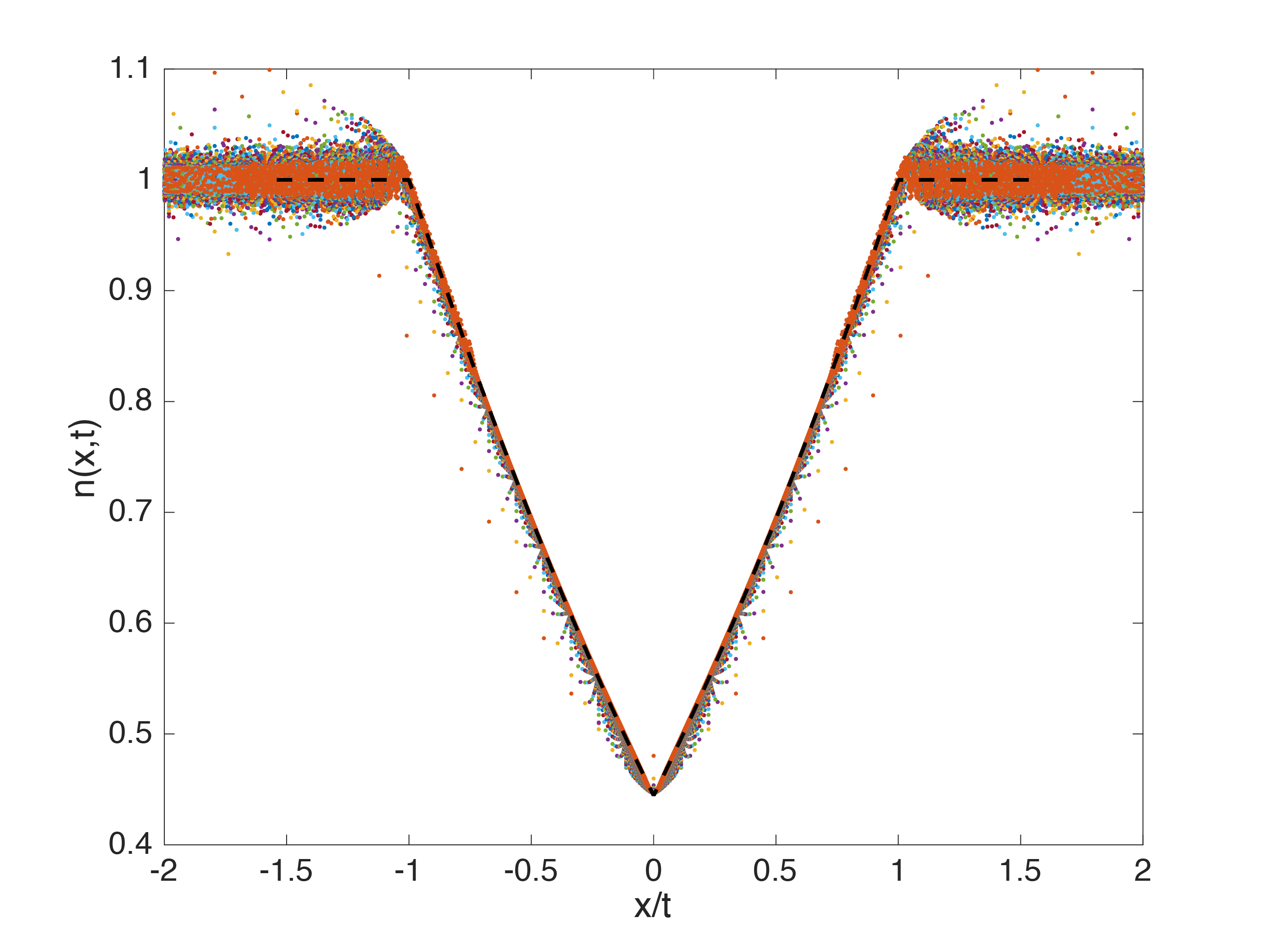}
\includegraphics[width=\columnwidth]{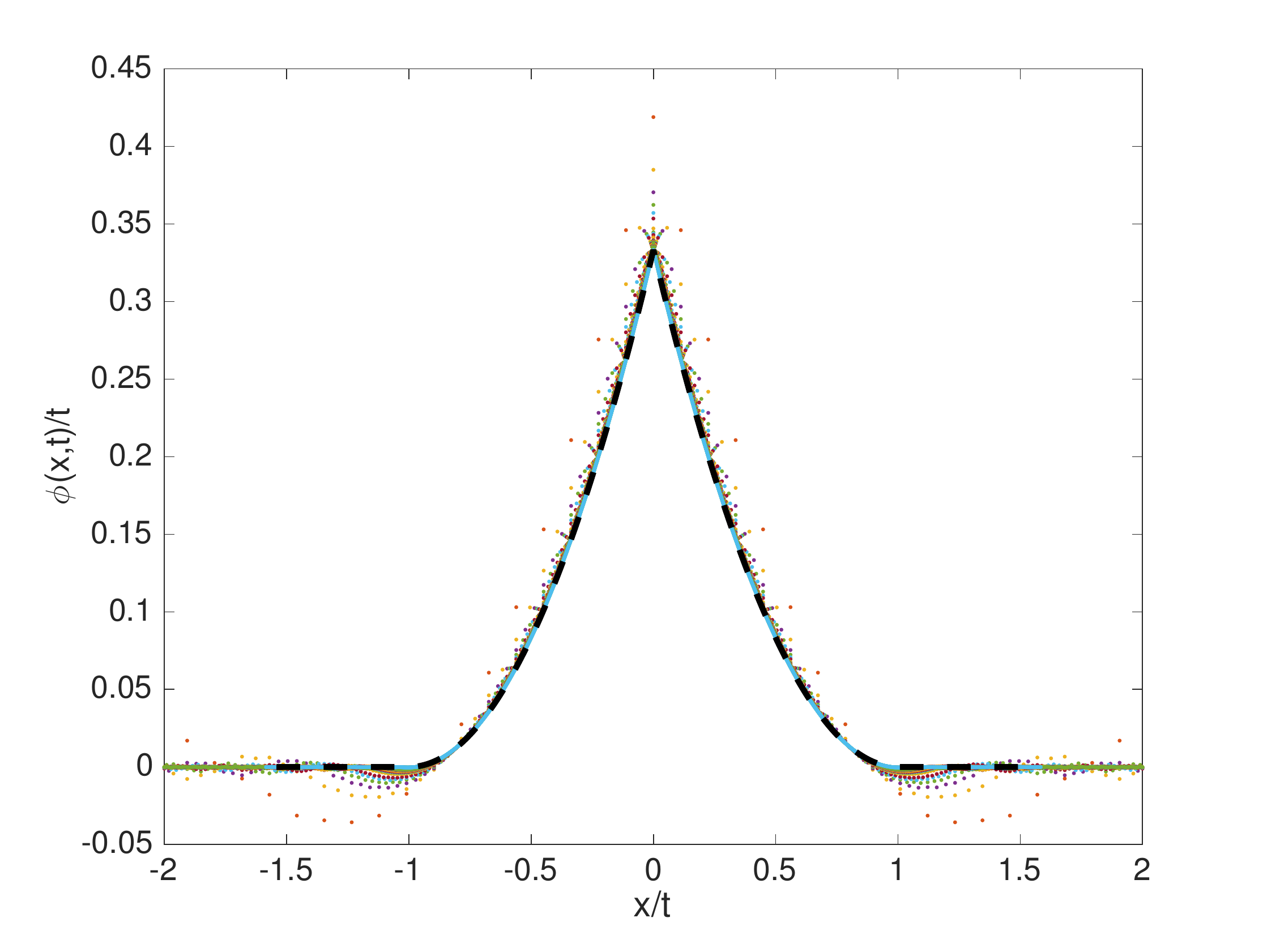}
\caption{{\bf Critical condensate} Critical condensate density and phase. The results show collapsed data for times $t=0$ to $t=500$. The black dashed line shows the analytic result. }
\label{fig:criticalscaling}
\end{figure}

So far, we have not discussed what the critical dissipation strength exactly is. Neither of the two stationary solutions allow us to uniquely fix the value of the critical point since they are stationary solutions for \emph{any} remaining condensate density. To do so, we need to take a closer look into the critical state of the system. Exactly at the critical loss it would take infinitely long to develop a homogenous background density and so the system is stuck in the transient state. Recall from Fig.~\ref{fig:metric} that this transient regime is accompanied with a quadratic onset of the phase, such that the velocity linearly interpolates between 0 at the causal horizon and $\gamma$ inside the depleted region. Since the system is stuck in this transient regime, let's assume that the phase $S$ at the critical point depends quadratically on $x$.  Imposing the constraint that the velocity vanishes when $x=c_0t$ and imposing that the density $n(x)$ equals the unperturbed density at the same point we find a unique solution the GPE within Thomas-Fermi approximation:
\begin{equation}
n=\frac{4n_0}{9}\left(\frac{|x|}{2c_0t}+1\right)^2, \quad S=\frac{c_0^2 t}{3} \left( \frac{|x|}{c_0t}-1 \right)^2-c_0^2t.
\end{equation}
The solution is completely fixed by the boundary conditions at $x=c_0t$, hence there is no more free parameter to fix the cusp in the phase in the origin. The latter uniquely fixes the critical point as it implies that this can only be solution of the dissipative GPE for $\gamma=2c_0/3$. This is in perfect agreement with the numerical results in Fig.~\ref{fig:dens_phase}. Moreover, in Fig.~\ref{fig:criticalscaling} we compare the analytical solutions of the critical density and phase with the numerical results. We observe almost perfect collapse on the analytic result. Corroborating the existence of a dissipative phase transition at $\gamma=2c_0/3$.

\section*{Fluctuations}
The classical condensate undergoes some interesting dynamics due to the local loss, but what about the quantum fluctuations? Is there anything interesting in the noise correlations of system? 

In the semi-classical limit, where we neglect the effects of the quantum pressure, small phase fluctuations $\delta S$ around the stationary state are approximately described by a relativistic wave equation~\cite{garay,carusotto02}: 
\begin{equation}
\frac{1}{\sqrt{-g}} \partial_\mu \sqrt{-g} g^{\mu \nu} \partial_\nu \, \delta S=0,
\end{equation}
where 
\begin{equation}
g^{\mu \nu}= \begin{bmatrix}
    1 &  v \\
    v & v^2-c^2
\end{bmatrix},
\end{equation}
$g=\det g_{\mu \nu}$, $v=\partial_x S/m$ and $c=gn/m$.  This suggest that nothing special happens below the critical loss rate and a horizon will be formed at the critical loss from which one could expect Hawking emission. However, a critical analysis shows that neither regimes can be treated in the hydrodynamic approximation. In the subcritical region, the condensate is homogenous and flow is constant on either side of the drain but is discontinuous at the drain. The effect of the latter is beyond the acoustic metric approximation. Also, above the critical loss the hydrodynamic description breaks down. Eventhough the horizon could be very far away from the drain, we know from~\eqref{eq:soliton} that it's actually always at the red shifted healing length. The latter once more implies that we can not neglect uv-properties of the system.

 \begin{figure}[h]
\includegraphics[width=\columnwidth]{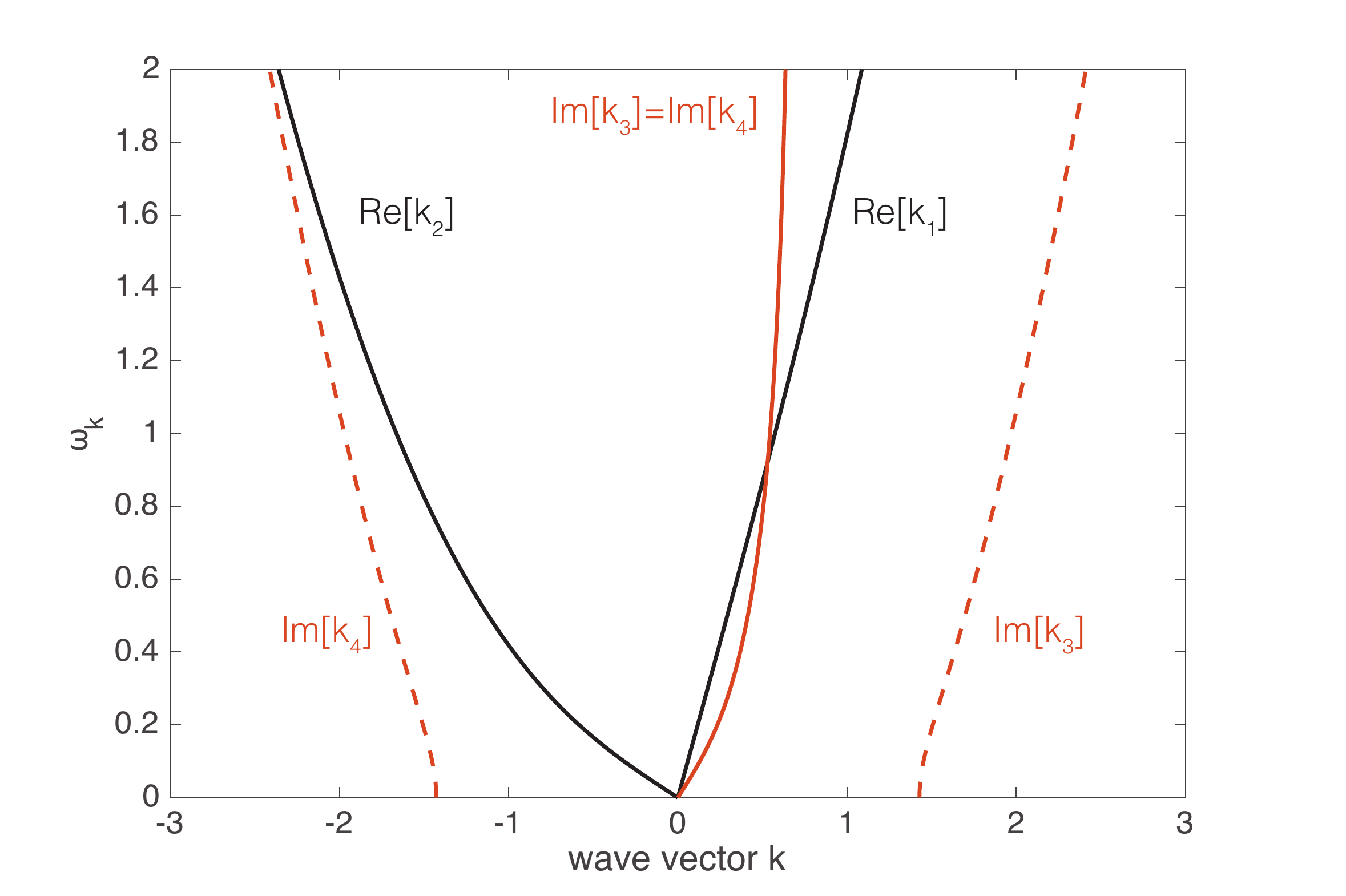}
\caption{{\bf Bogoliubov dispersion} The black lines show the usual positive norm Bogoliubov modes. Both have completely real wave vectors at positive energy. The red lines show the negative Bogoliubov norm modes. The latter have complex wave vectors, the full line shows the real part of $k$ while the dashed line shows the imaginary part. Figure shows the dispersion on the left of the drain for at $v=0.7c$.}
\label{fig:dispersion}
\end{figure}

As long as the system is dilute, small fluctuations $\chi$ around the condensate $\psi_0$ can still be treated in Bogoliubov approximation. From this point on, let us work in units of $\hbar=m=gn=1$, such that the Bogoliubov equations become
\begin{equation}
i\hbar \partial_t \begin{pmatrix}
    \chi \\
    \chi^\ast
\end{pmatrix}=
\begin{pmatrix}
    H_0 &  (\psi_0)^2 \\
    -(\psi_0^\ast)^2& -H_0^\dagger
\end{pmatrix}\begin{pmatrix}
    \chi \\
    \chi^\ast
\end{pmatrix} +\delta(x) \begin{pmatrix}
    \eta(t) \\
    \eta^\ast(t)
\end{pmatrix},
\end{equation}
where 
\begin{equation*}
H_0=-\frac{1}{2} \partial^2_x +2 |\psi_0|^2-\mu-i\gamma \delta(x) 
\end{equation*}
Far from the drain, the eigenvectors of the Bogoliubov problem can be found exactly; both below and above the critical loss. Consequently, we can analyze the dynamics of the fluctuations by constructing the scattering solutions. The aymptotic stationary solutions are of the form
\begin{equation}
\chi(x,t)=e^{-i\omega t} u(x)+e^{i\omega t} v^\ast(x),
\end{equation}
where we restrict to positive frequencies. The latter is not crucial but avoids double counting the same mode. Let's first focuss on the behavior below the critical loss. In that case the system is homogenous and the eigenvectors are simply plane waves, such that we get the usual Bogoliubov dispersion:
\begin{equation}
(\omega-vk)^2=\left( \frac{k^2}{2}+1\right)^2-1.
\label{eq:dispersion}
\end{equation}
Note that the set of Bogoliubov equations are forth order secular equations in $k$. We thus need 4 independent solutions at every frequency in order to satisfy the boundary conditions. Apart from the two propagating modes there are two additional complex solutions to eq.~\eqref{eq:dispersion}, see Fig.~\ref{fig:dispersion}. These complex solutions simply represent evanescent solutions to the Bogoliubov equations. Moreover, these solutions have negative Bogoliubov norm, i.e. $|u|^2-|v|^2=-1$ such that upon quantization creation and annihilation operators are switched. Consequently, any process that can couple the propagating positive norm solution and the localized negative norm solution will result in spontaneous creation of particles out of the vacuum. 

 \begin{figure}[h]
\includegraphics[width=\columnwidth]{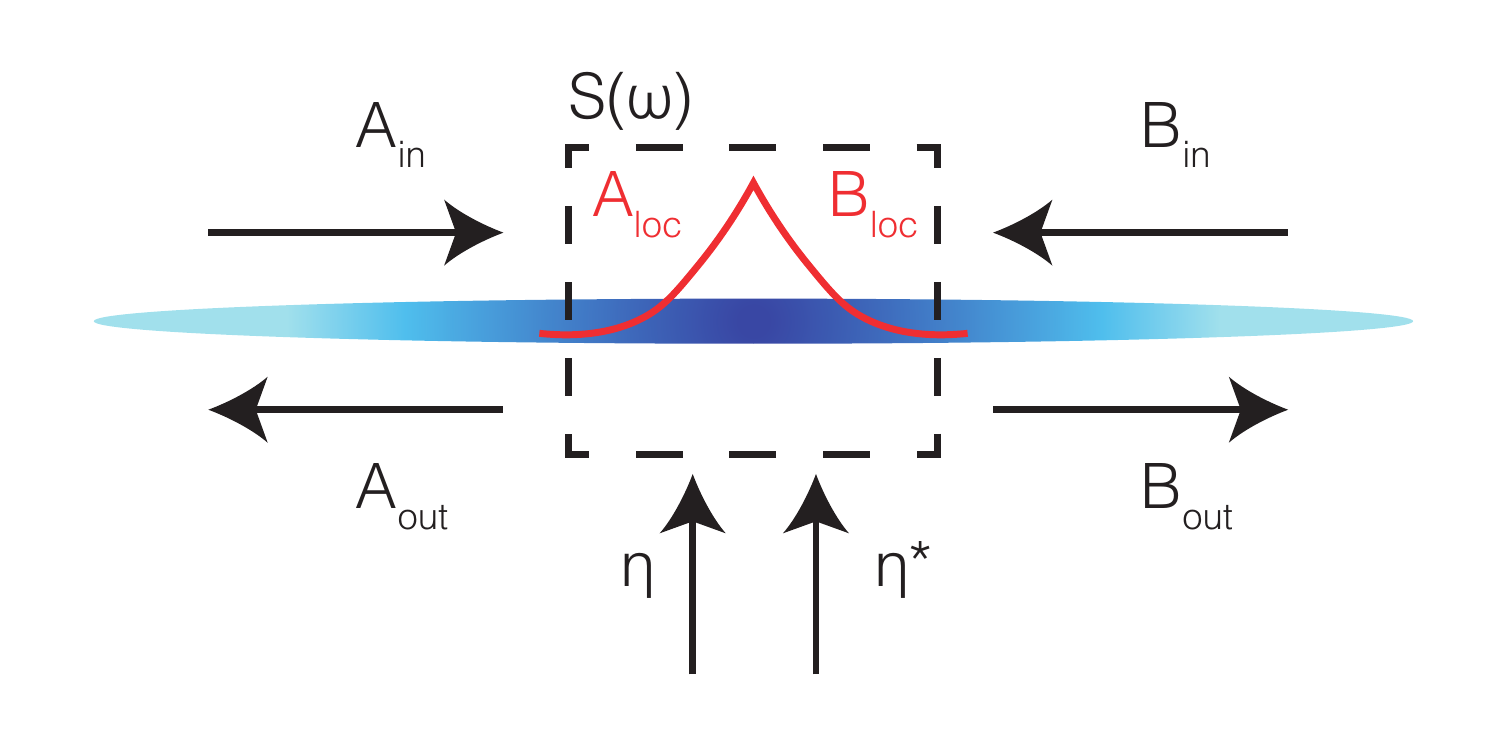}
\caption{{\bf Scattering modes} The scattering problem involves a set of 8 modes. Apart from the usual in and out propagating modes on the left and right, denotes by $A_{in/out}$  and $B_{in/out}$ respectively. There are two evanescent modes, $A_{loc}$ and $B_{loc}$ and there is the vacuum noise $\eta$ caused by the dissipation.  }
\label{fig:scatteringSetting}
\end{figure}

In our dissipative problem there is a second process that can lead to particle generation. Apart from mixing positive and negative norm Bogoliubov modes of the condensate, quantum noise from the dissipation also enters the system, see Fig.~\ref{fig:scatteringSetting}. On the quantum level the process is similar to the mixing between the evanescent and propagating modes. Just as the latter, this process involves two mode squeezing. Since the dissipation removes atoms, its quantum vacuum is simply the state with no atoms inside. However, the quantum vacuum of the condensate is the state with no Bogoliubov excitations, hence annihilating an atom involves squeezing the condensate-reservoir state. The main difference with the former process is that, unlike the localized modes in the condensate, we don't have access to the reservoir and we can only measure its effect on the condensate.

 \begin{figure}[h]
\includegraphics[width=\columnwidth]{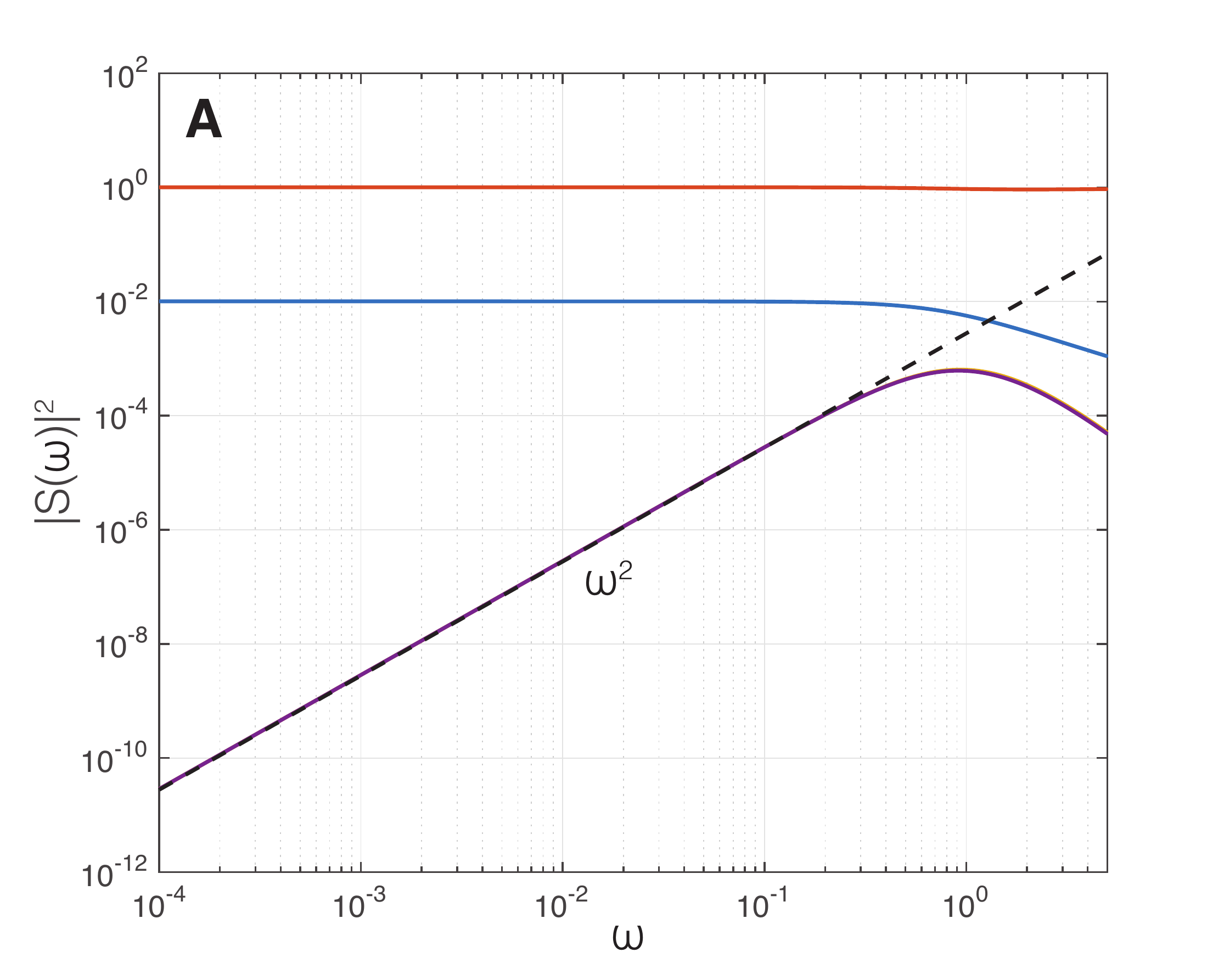}
\includegraphics[width=\columnwidth]{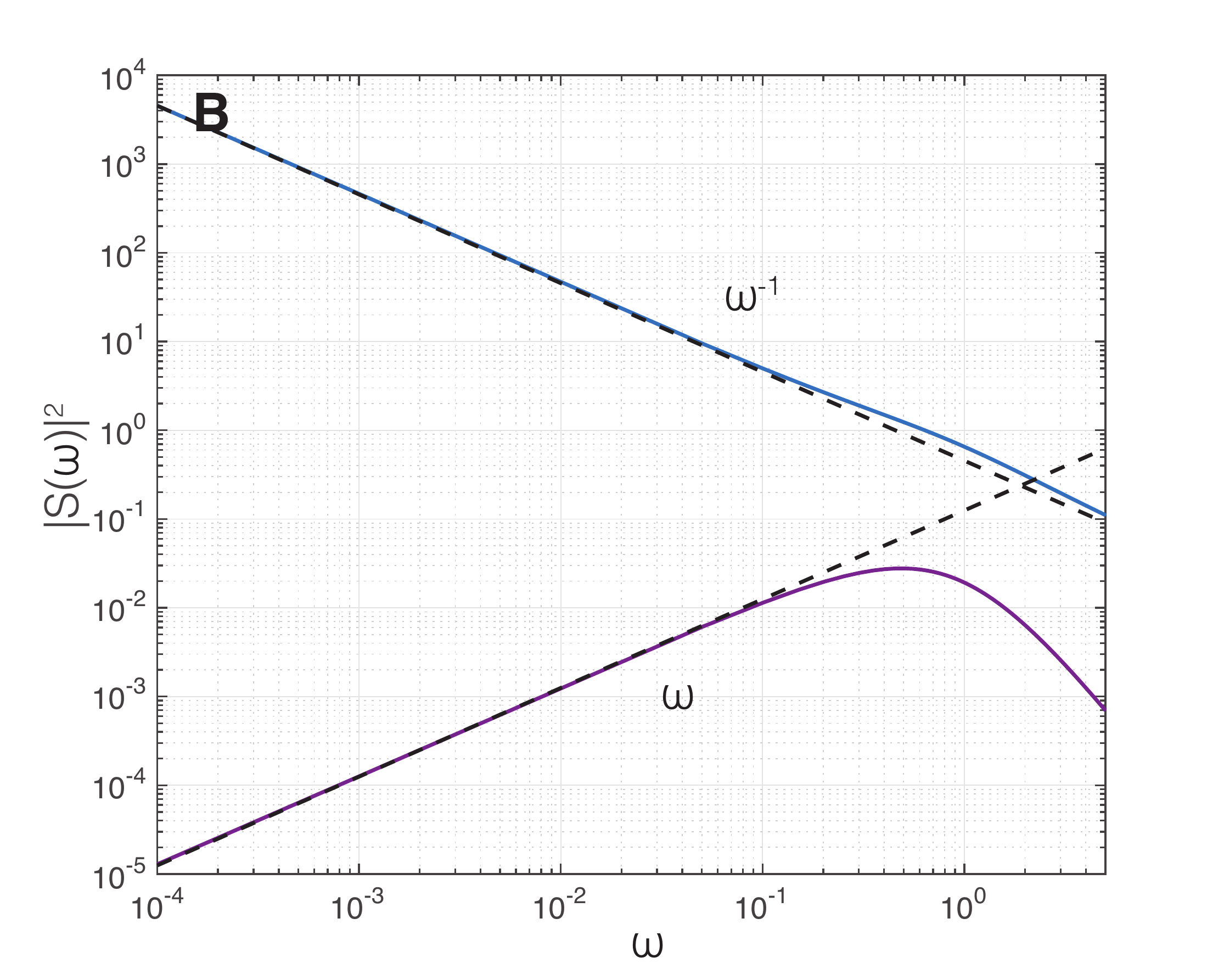}
\caption{{\bf Scattering solutions} Different components of the S-matrix for relatively weak dissipation of $\gamma=0.1$ are shown. Panel A shows that the incident wave is almost completely transmitted (red line), small part is reflected (blue line) and a small part scatters into the evanescent modes (purple line). Panel B shows the S-matrix elements for the coupling to the quantum noise from the dissipation. The blue line shows the intensity in the outgoing channel and the purple line shows the coupling to the localized mode. }
\label{fig:scatteringSol}
\end{figure}

Figure~\ref{fig:scatteringSol}  shows the solution the scattering problem at weak dissipation, much below the critical loss. As expected, most of the incoming wave is transmitted but $O(\gamma^2)$ is reflected back. At zero frequency, there is is no coupling to the evanescent modes as the intensity in that mode increases quadratically with frequency. The latter was recently observed in Ref.~\cite{curtis}; the vanishing matrix elements was linked to the absence of static black hole hair~\cite{kai}.

As shown in panel B of Fig.~\ref{fig:scatteringSol}, the dominant process at low frequency is actually outcoupling of the dissipative noise $\eta$ into the outgoing and evanescent modes in the condensate. Let us thus first restrict our attention to the noise modes. At low frequency we can find the asymptotic values of the S-matrix analytically, moreover we can just focuss on one of the sides since everything is symmetric. The outgoing modes have:
\begin{equation}
S_{out}(\omega)= \sqrt{\frac{1}{2(1+v)\omega}} \left( e^{-i\phi}\eta+e^{i\phi}\eta^\ast\right),
\label{eq:Sout}
\end{equation}
where $\cos \phi=v$. Note that this implies that the number of phonons in the outgoing mode is
\begin{equation}
p_{out}(\omega)=|S_{out}(\omega)|^2=\frac{1}{2(1+v)\omega}  \left| e^{-i\phi}\eta+e^{i\phi}\eta^\ast\right|^2,
\end{equation}
averaging over the noise yields:
\begin{equation}
p_{out}(\omega)=\frac{v}{1+v} \frac{1}{\omega}.
\end{equation}
We thus get classical equipartition and the temperature is $k_bT=v/(c+v) \mu$. Remarkably, we still get thermal behavior just like Hawking radiation, eventhough there is no horizon in the acoustic metric.

\begin{figure}[h]
\includegraphics[width=\columnwidth]{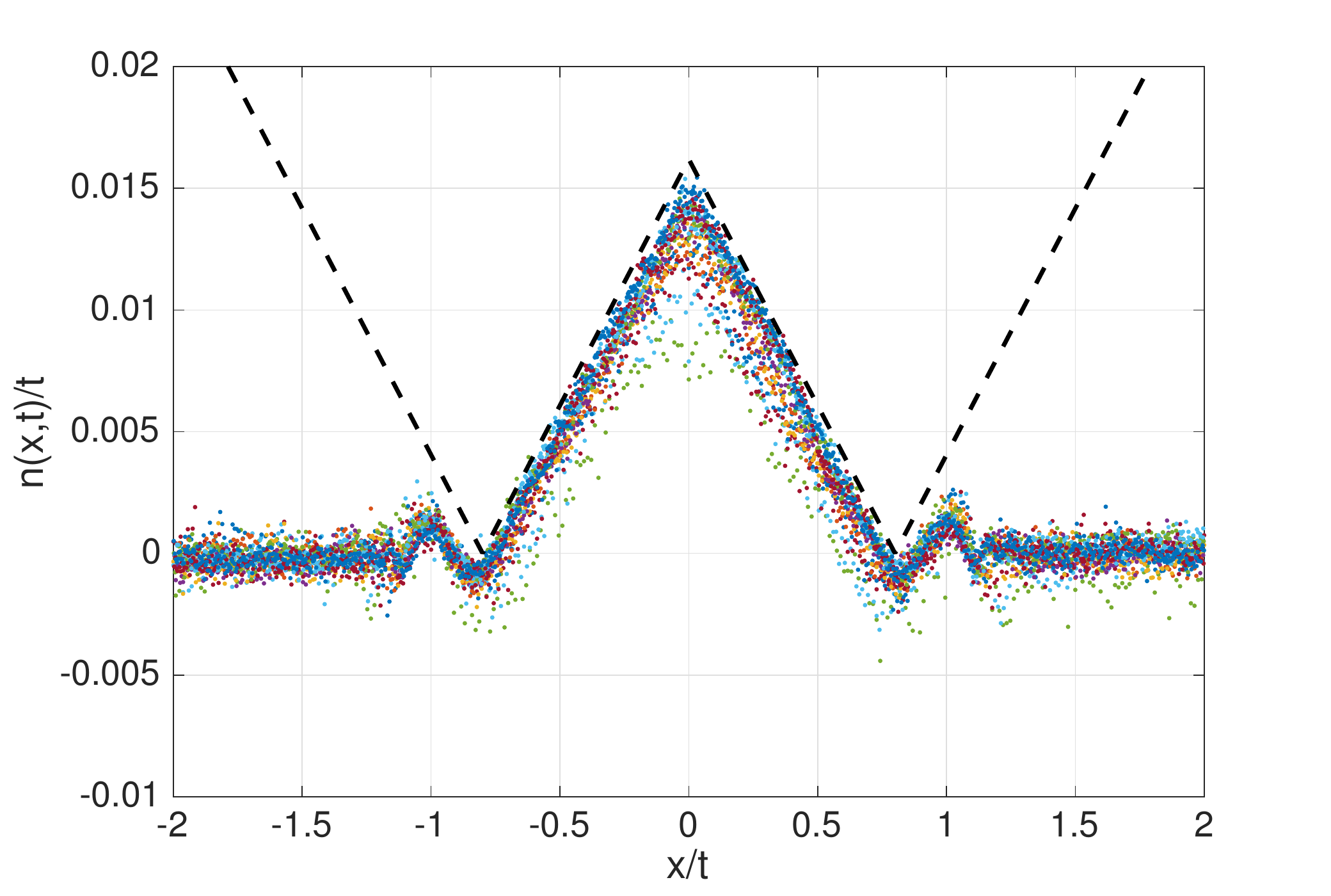}
\caption{{\bf Condensate fluctuations} Plot shows the scaled density fluctuations for TWA simulations at $\gamma=0.1$. The dashed line shows the analytic result based on the scattering solutions of the Bogoliubov theory. }
\label{fig:g1}
\end{figure}

The thermal occupation of phonons can be seen in the fluctuations of the atomic condensate. Indeed, at low energy the Bogoliubov coefficients are
\begin{equation}
u_{out}\approx -v_{out}\approx \sqrt{\frac{1-v}{2\omega}},
\label{eq:uvout}
\end{equation}
while the wave vector of the outgoing modes is simply $k_{out}=\omega/(1-v)$. Consequently, after averaging over the noise and integrating over all frequencies we obtain the following expression for the condensate fluctuations:
\begin{equation}
n_{out}(x,t)=\left< \chi^\dagger(x,t) \chi(x,t) \right>=\frac{1}{2}\frac{v}{1+v} ||x|-(1-v)t|.
\end{equation}
The temperature is thus directly related spatial derivative of the condensate fluctuations. Unlike the phonon occupation, the latter can directly be extracted from TWA simulations. In fig.~\ref{fig:g1}, we show excellent agreement between the simulations and the Bogoliubov scattering theory. 

Single particle fluctuations are completely dominated by fluctuations induced by coupling to the external reservoir. However, since single particles are annihilated by the dissipation, those fluctuations are independent to order $\gamma$. Correlations between successive events are only of order $\gamma^2$. Consequently, almost all those fluctuations can be removed by looking at density-density correlation:
\begin{equation}
g^{(2)}(x,x')= \frac{\left< \psi^\dagger(x)\psi^\dagger(x')\psi(x')\psi(x) \right>_c}{\left<\psi^\dagger(x)\psi(x)\right>\left<\psi^\dagger(x')\psi(x')\right>}
\end{equation}
Indeed, computing the density-density correlations between the outgoing modes using~\eqref{eq:Sout} immediately shows all terms of order $\eta^2$ cancel. Contributions to $g^{(2)}$ are only due to genuine two-particle noise correlations. Not only are those of $O(\gamma^2)$, they are further suppressed by a factor $1/n$ which makes them completely irrelevant. 

\begin{figure}[h]
\includegraphics[width=\columnwidth]{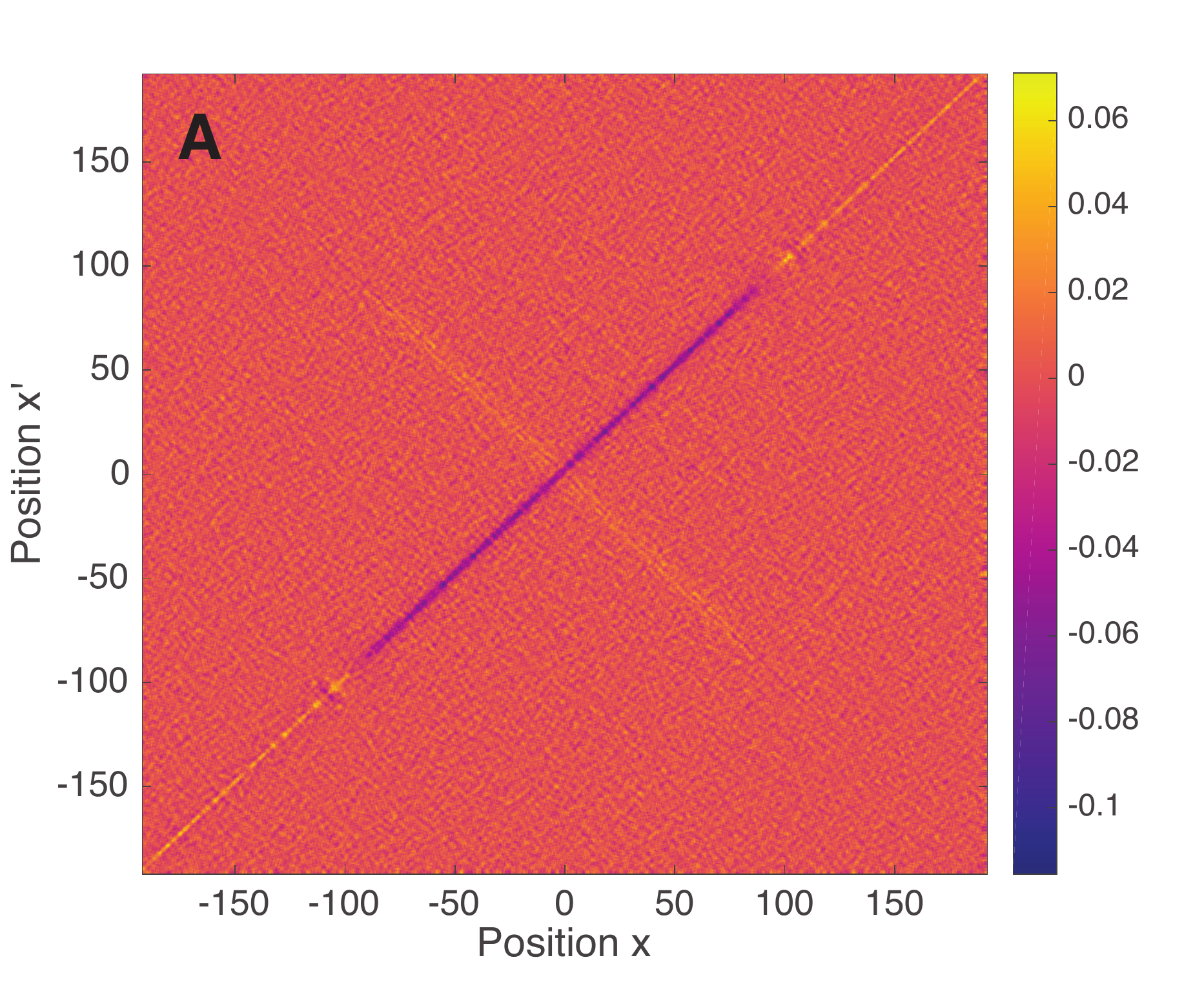}
\includegraphics[width=\columnwidth]{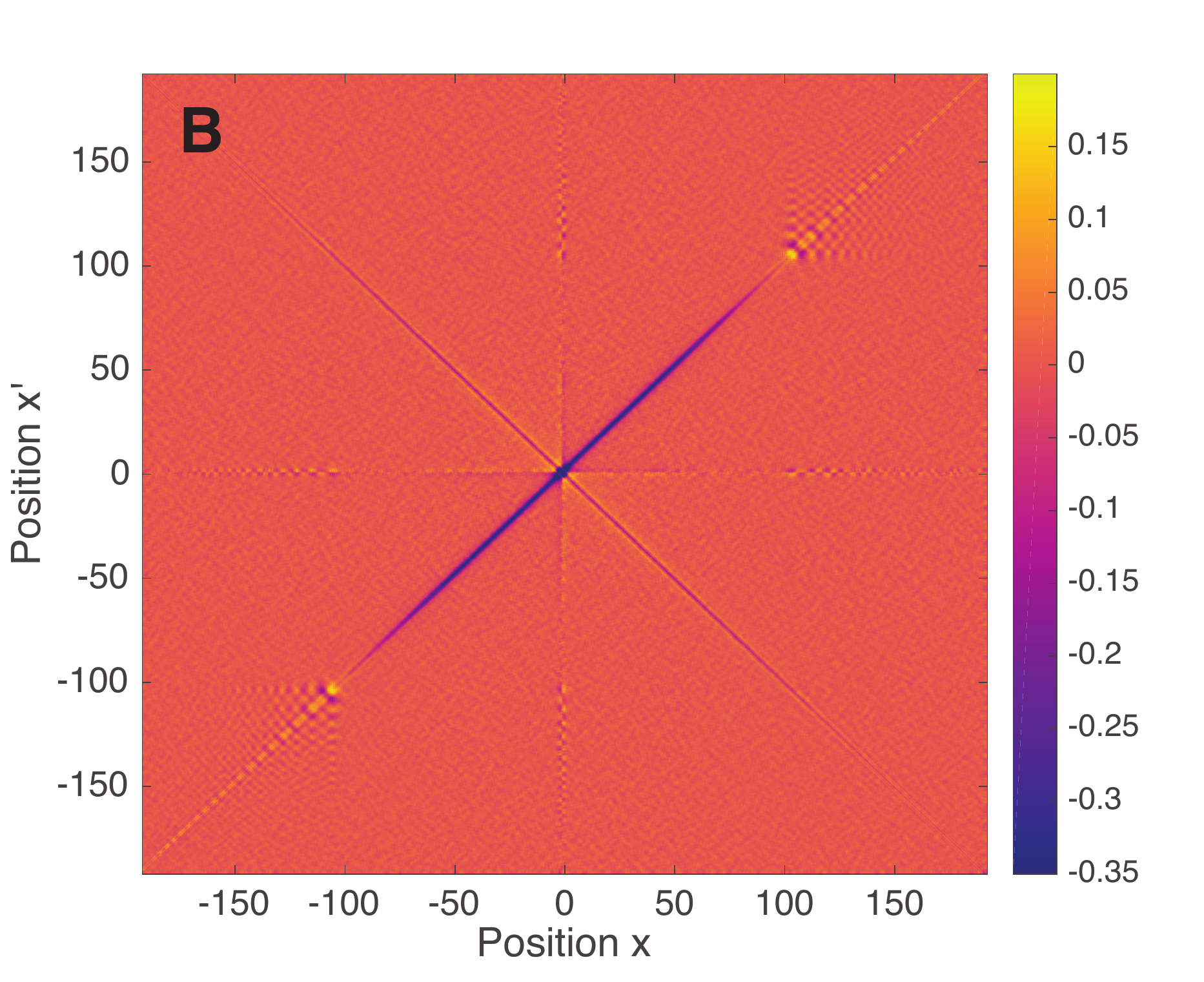}
\caption{{\bf Density-Density correlation}  After connecting the condensate to the reservoir correlations develop in the system. Panel~{\bf A} shows $g^{(2)}(x,x')$  for $\gamma=0.1$ after $t=100$.  Panel~{\bf B} shows $g^{(2)}(x,x')$  for $\gamma=2$ after $t=100$. Both panels only shows the difference between time-dependent correlations and the initial correlations in the system. Moreover, they are normalized with respect to the initial diagonal density correlation, this way the results is of order $\gamma$. Data is computed from $10^4$ TWA realizations.  }
\label{fig:g2}
\end{figure}

Two particle correlations are therefore induced by coherent scattering processes between atoms in the condensate. As shown in panel A of Fig.~\ref{fig:scatteringSol}, scattering to the localized mode is largely suppressed at low frequency. At zero frequency, the reflection and transmission actually become 
\begin{equation}
r=-\frac{v}{\sqrt{1-v^2}}, \quad t=\frac{1}{\sqrt{1-v^2}}.
\end{equation}
In order to look at two-particle fluctuations we must simply check what happens to the initial $g^{(2)}(x,x')$ because of the scattering in the origin. Firstly, recall that the isolated condensate has particle anti-bunching because of interactions. The latter can also be thought of as a positive correlation between holes. Particles are being annihilated and so our scattering process is non-unitary. For weak loss $O(v)$ particles are annihilated from either side or an extra amount of $v$ holes is created. Therefore we expected amplification of $g^{(2)}(x,x')$ close to the diagonal by a factor of $v$. The only other process that contributes to the density-density correlation is the correlation between the back-reflected and the transmitted wave. Those should have a correlation which is proportional to $r\times t$ and should appear on the anti-diagonal $x=-x'$. Because there is a phase-shift op $\pi$ upon reflection, those processes should be positively correlated. Moreover, the signal from the left and the right do not add up as the back-scattered waves travel in opposite direction. Therefore, we expect the signal to be about half as strong as the extra anti-bunching on the diagonal. 
Numerically we extract $g^{(2)}$ from a microscopic simulation using TWA. The result's shown in panel A of Fig.~\ref{fig:g2} agree well with the expectations from scattering theory.

\begin{figure}[h]
\includegraphics[width=\columnwidth]{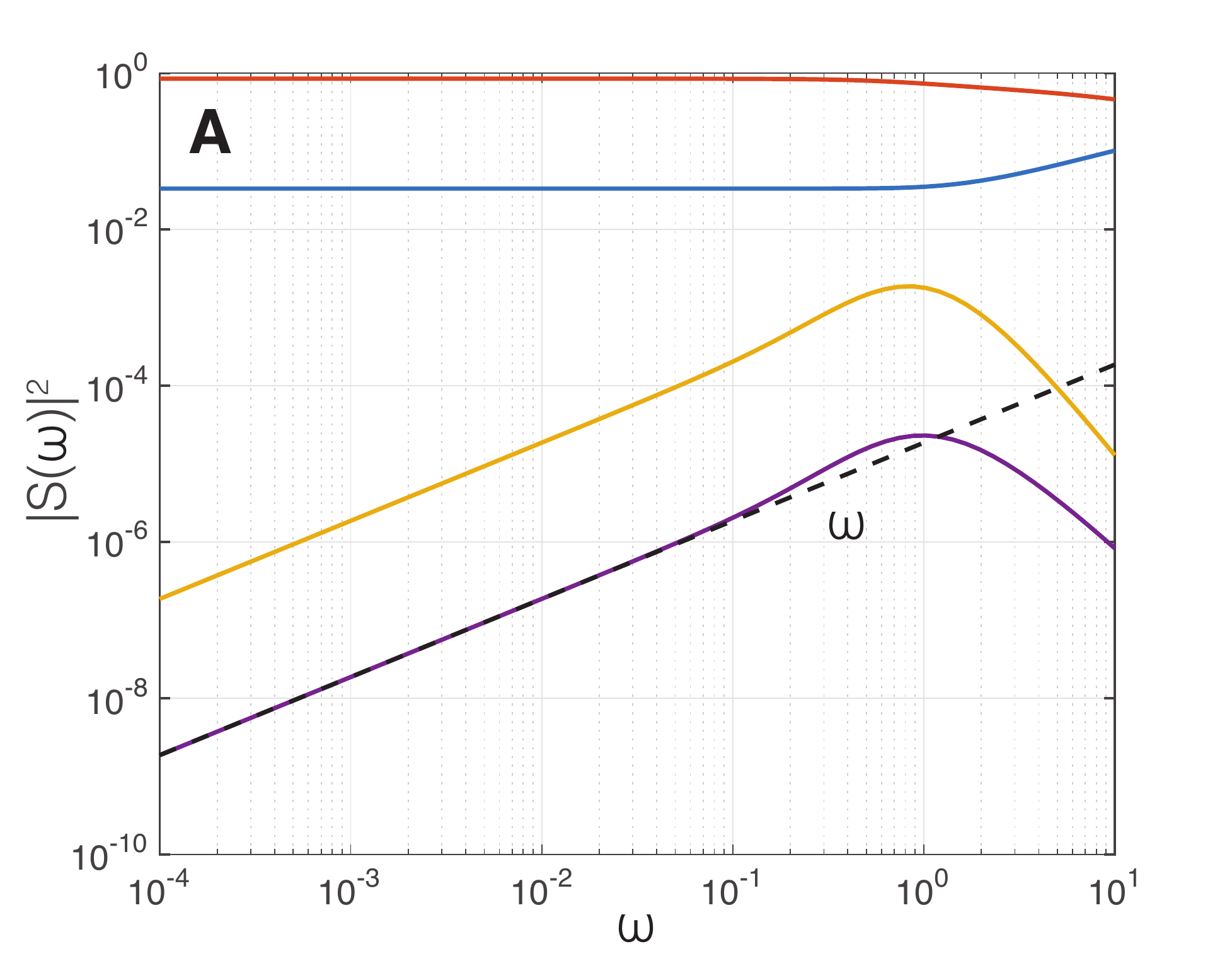}
\includegraphics[width=\columnwidth]{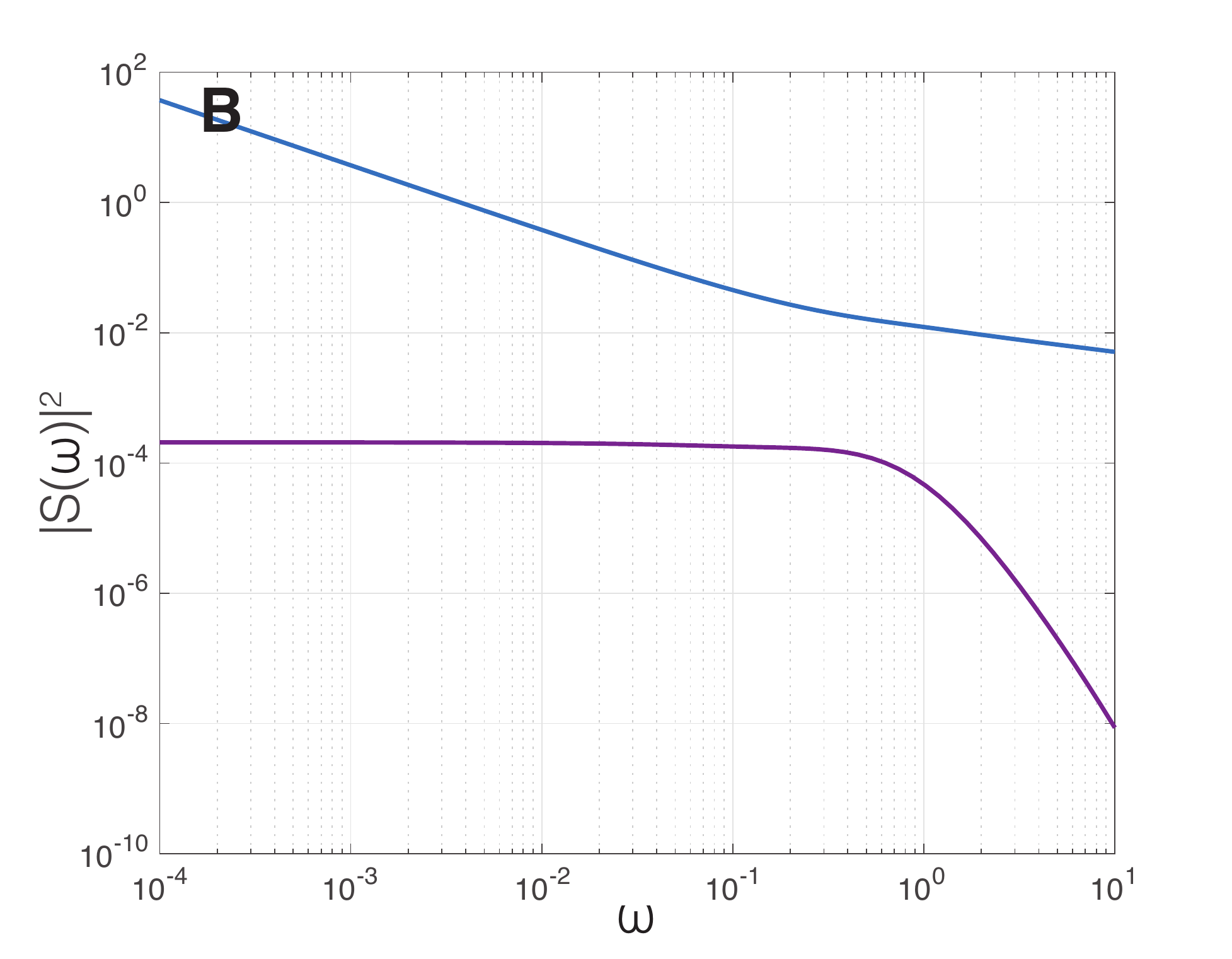}
\caption{{\bf Scattering solutions} Different components of the S-matrix for strong dissipation of $\gamma=10$ are shown. Panel A shows that the incident wave is mostly transmitted (red line), small part is reflected (blue line) and a small part scatters into the evanescent modes (purple line for reflected and yellow for transmitted). Panel B shows the S-matrix elements for the coupling to the quantum noise from the dissipation. The blue line shows the intensity in the outgoing channel and the purple line shows the coupling to the localized mode. }
\label{fig:scatteringSoliton}
\end{figure}

\begin{figure}[h]
\includegraphics[width=\columnwidth]{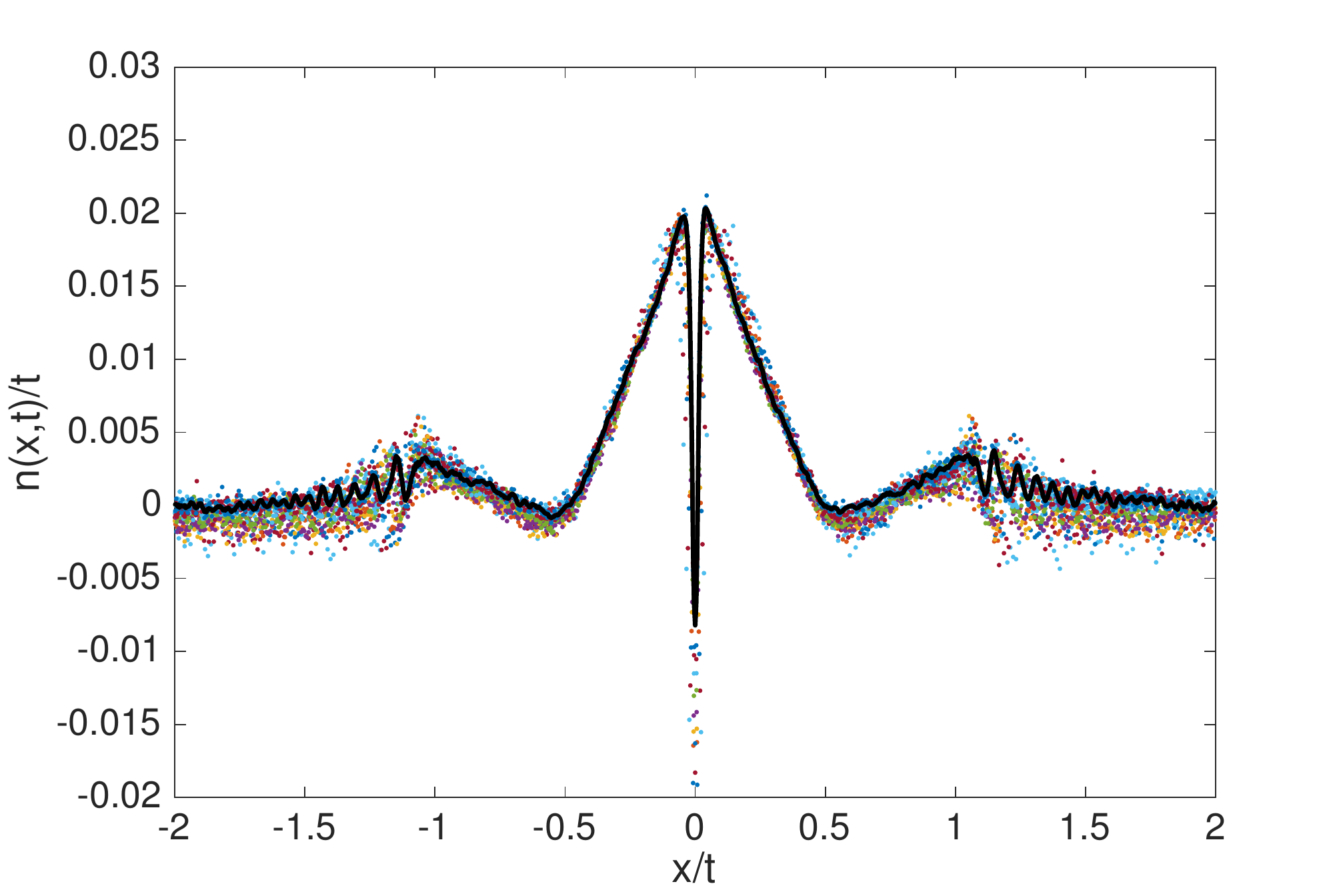}
\caption{{\bf Condensate fluctuations} Plot shows the scaled density fluctuations for TWA simulations at $\gamma=2$. Compared to~Fig.~\ref{fig:g1} a significant dip is present around the drain caused by coupling to the evanescent modes}
\label{fig:g1bh}
\end{figure}

The same procedure can be repeated to construct the scattering solution above the critical loss. The exact Bogoliubov modes for the inhomogeneous condensate~\eqref{eq:soliton} have been constructed in Ref.~\cite{walczak}. Remarkably, eventhough there seems to be an acoustic horizon, the dispersion of the eigenmodes is exactly the same as~\eqref{eq:dispersion}. There are no supersonic modes and the only negative norm modes are the same evanescent modes as before. However, as show in Fig.~\ref{fig:scatteringSoliton} for $\gamma=10$, the coupling to this evanescent modes has drastically changed behavior. 
The S-matrix elements have picked up an extra factor of $1/\sqrt{\omega}$, making them relevant in the IR. For the single particle fluctuations it induces an additional dip in the condensate fluctuations around the drain, see Fig.~\ref{fig:g1bh}.
The density-density correlations are shown for $\gamma=2$ in panel B of Fig.~\ref{fig:g2}. We get the same additional anti-bunching on the diagonal but the positive correlation between reflected and transmitted particles has become negative. The latter is consistent with the absence of a phase shift in the scattering problem. More strikingly, correlations between scattered and localized modes shown up. While the correlations between reflected and localized waves is negative, it is positive between transmitted and localized modes.  

\section*{2 drains and lasing}
\begin{figure}[h]
\includegraphics[width=\columnwidth]{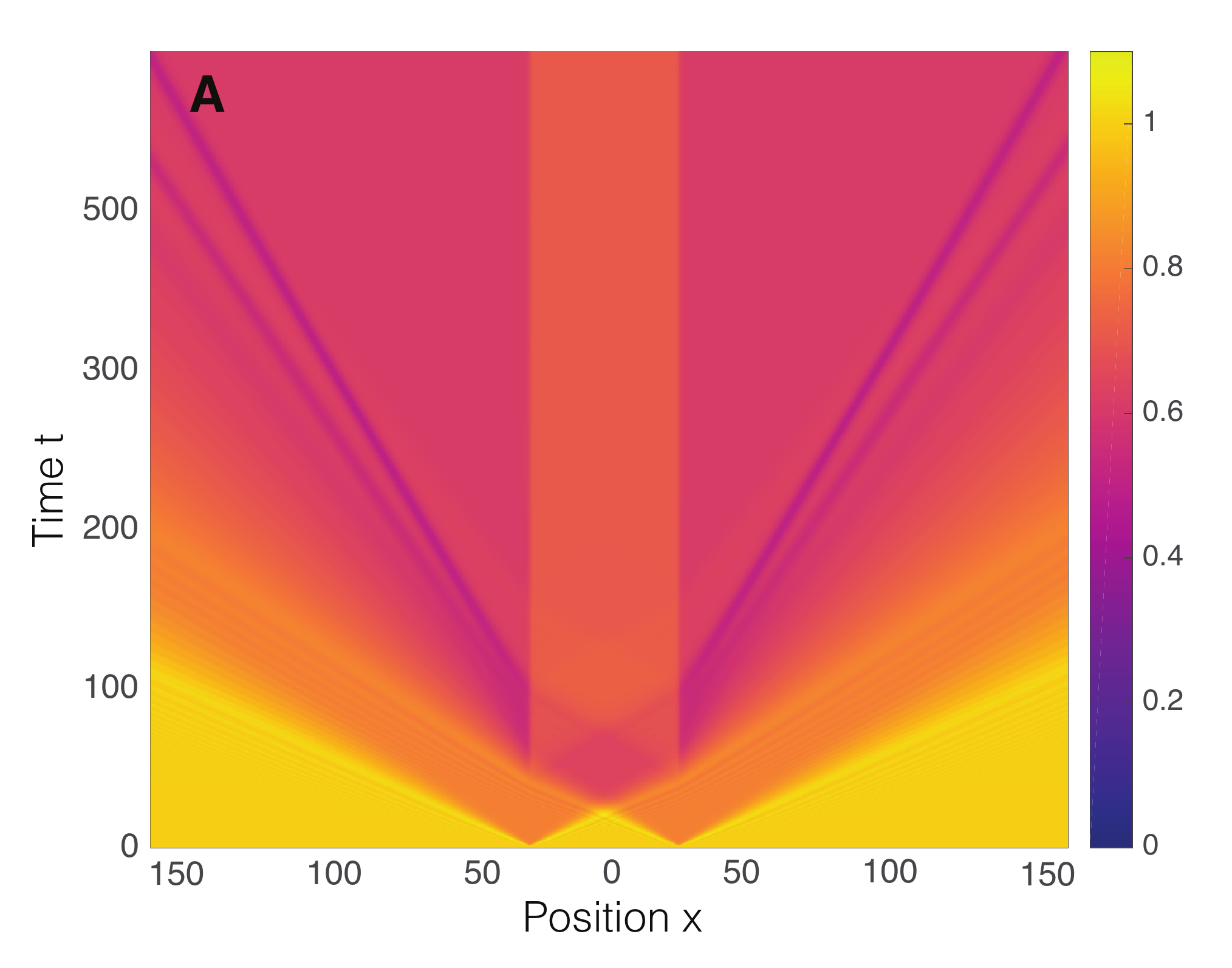}
\includegraphics[width=\columnwidth]{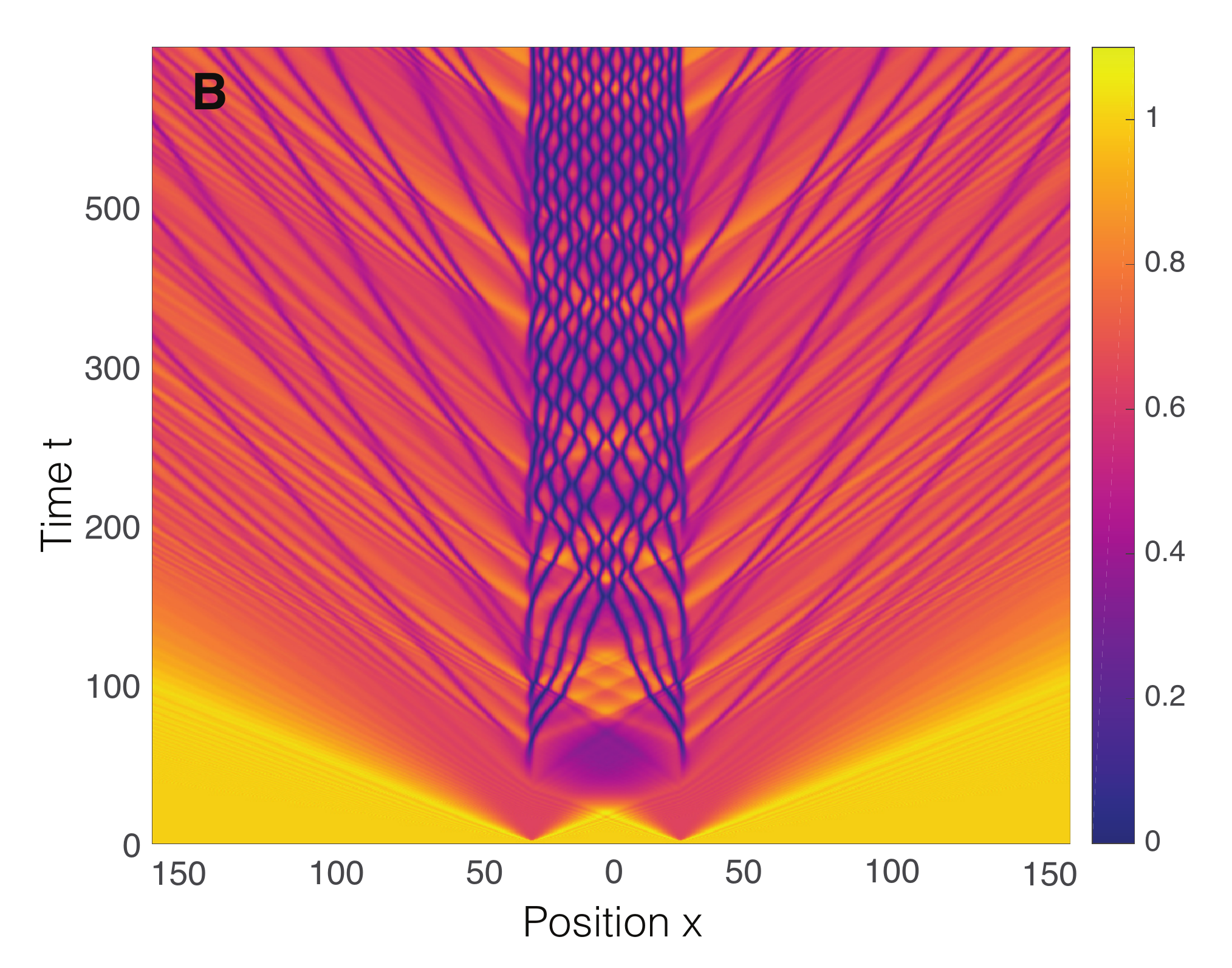}
\caption{{\bf Two drain density} The density is shown for two drains located $L=60\xi$ away from each other. At weak loss, the result is close to a superposition of two individual drains; panel {\bf A} for $\gamma=0.1$. However, when the loss exceeds a critical value such that the combined flow generates a horizon, the result is drastically different. As shown in panel~{\bf B} for $\gamma=0.4$, solitons are generated at the drains and fill up the region in between the horizons. }
\label{fig:2drain}
\end{figure}

\begin{figure}[h]
\includegraphics[width=\columnwidth]{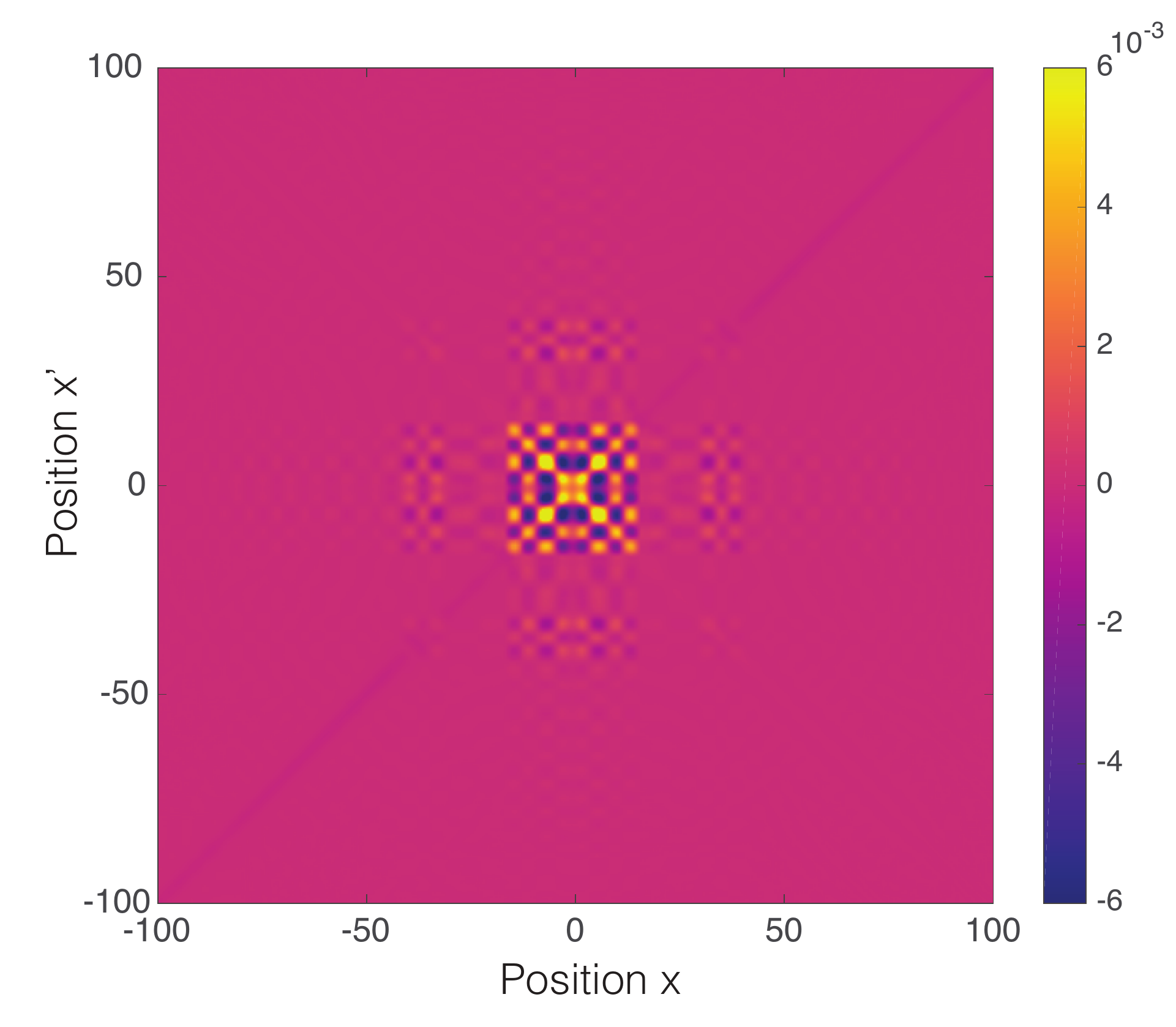}
\caption{{\bf Correlations hawking laser} Strong correlations develop in the Hawking laser. Figure shows $g^{2}$ for two drains located at $L=30\xi$. Data is computed from $10^4$ TWA realizations. }
\label{fig:g2_laser}
\end{figure}

While the density-density correlation shows clear correlations between localized and emitted particles in the Hawking regime, one might still be concerned with the coherent nature of the process. If coherent, two black holes should be able to stimulate emission in each-other, effectively resulting in a \emph{black hole laser}. The lasing instability was first experimentally shown in experiment~\citep{Steinhauer02}. By inducing (incoherent) dissipation at two spatially separated points, we can induce two independent flows. However, once the two drains are in each other causal region, we expect the flow in between the two drains to stop whereas the flow from the outside should be about twice as large as that of an individual drain. This implies the threshold for forming a horizon is drastically reduced. Moreover, this asymmetry in the flow around every drain allows for soliton shedding~\citep{vincent}. As shown in Ref.~\citep{florent}, solitons are shedded in the subsonic side; in the case in between the two drains (see Fig.~\ref{fig:2drain}). However, for a fixed distance between the drains, the system can only accommodate a finite amount of solitons such that the process saturates after some time. The resulting density-density correlation is shown in Fig.~\ref{fig:g2_laser} in the lasing regime. Strong correlations develop in the subsonic region in between the drains. A detailed numerical study of soliton emission and its relation to Hawking lasing was recently performed in Ref.~\citep{carusotto03}. The present configuration is one of the simplest multi-drain configurations to consider as the drains are completely balanced. Interesting spatio-temporal order seems to arises in situations in which the drains are unbalanced. The behavior of the latter is poorly understood and warrants further research.

\section*{Experimental considerations}
In this work we presented a detailed analysis of the evolution of a dissipative BEC into a NESS , including the formation of non-equilibrium correlations in the system. We conclude the manuscript with a brief discussion on some experimental considerations. To observe the thermal radiation of Bogoliubov phonons directly, the condensate must be cooled below the Hawking temperature before the dissipation is switched on. The latter is always smaller than the chemical potential but for realistic parameters of the dissipation it is of the same order. Cooling to temperatures below the chemical potential in a 1D degenerate quantum gas is challenging, as standard evaportive cooling becomes inefficient. However, temperatures as low as $0.25\mu$ have recently been demonstrated by coherently outcoupling atoms using a weak rf-drive~\citep{rauer}. For sufficiently cold initial condensates, the Hawking temperature can be extracted from correlations in the density specle pattern after time-of-flight expansion~\citep{manz,imambekov} or by performing an interference experiment on two condensates. 

Density-density correlations can be extracted from direct measurements of the density. At best, those are shot-noise limited. As shown in Fig.~\ref{fig:g2}, the signal is a fraction of the initial $g^{(2)}(0,0)$ and the latter scales inversely with the dilution parameter $n\xi$. In a typical experiment it's reasonable to expect $n\xi\approx 10$. For $\gamma=0.1$, expected density-density correlations are therefore of $O(10^{-2})$; implying there is about 1 correlated fluctuation in every healing length. Since the shot noise scales with $\sqrt{n\xi}$, $O(100)$ measurements are required to get sufficient statistics to resolve the predicted correlation pattern. Alternatively, correlations in phase fluctuations provide complementary information to the density-density correlations and can be measured by interfering two condensates. 
The most challenging part of our proposal is the realization of controlled local loss. In particular, the loss has to be engineered on a scale comparible to the healing length. Loss applied over an extended region results in a lasing instability comparible to the 2 drain results shown above.

Finally, throughout this work we have focussed on the behavior of a BEC. However, most features are expected to be generic to a broad class of systems. Any system that, in the hydrodynamic limit, can be described by a relativistic wave equation and which has a violation of Lorentz-invariance at higher energy that allows for supersonic propagation of excitations is expected to undergo a similar instability towards a black hole state.

\begin{acknowledgments}
\emph{Acknowledgements.---} DS acknowledges support from the FWO as post-doctoral fellow of the Research Foundation -- Flanders. The authors acknowledge support from  
Harvard-MIT CUA, NSF Grant No. DMR-1308435, AFOSR-MURI Quantum Phases of Matter (grant FA9550-14-1-0035), AFOSR-MURI: Photonic Quantum Matter (award FA95501610323). The authors like to thank M. Wouters, H. Ott, W. Unruh, S. Fagnocchi, J. Schmiedmayer, J. Dalibard, I. Bloch for useful discussions.
\end{acknowledgments}

\bibliographystyle{apsrev4-1}
\bibliography{Blackhole.bib}

\end{document}